%% file: ITW25.tex
\documentclass[conference,letterpaper]{IEEEtran}
\usepackage[top=0.73in,bottom=1.05in,left=0.637in,right=0.64in]{geometry}
\IEEEoverridecommandlockouts

\usepackage{enumerate}
\usepackage[utf8]{inputenc} 
\usepackage[T1]{fontenc}
\usepackage{url}
\usepackage{ifthen}
\usepackage{tcolorbox}
\usepackage[noadjust]{cite}
\usepackage[cmex10]{amsmath} 
\usepackage{amssymb,amsfonts,amscd,amsbsy}
\usepackage{graphicx}
\usepackage{stmaryrd} % for \llbracket and \rrbracket
\usepackage[mathscr]{eucal}
\usepackage{mathtools}
\usepackage{bbm}
\usepackage{xcolor}
\usepackage{xfrac}
\usepackage{algorithm}
\usepackage{algpseudocode}
\usepackage{enumerate}
\usepackage{enumitem}
\usepackage{multirow}
\newtheorem{theorem}{Theorem}
\numberwithin{theorem}{section}
\newtheorem{corollary}[theorem]{Corollary}
\newtheorem{lemma}[theorem]{Lemma}
\newtheorem{definition}[theorem]{Definition}
\usepackage{etoolbox}
\newcommand{\overbar}[1]{\mkern 1.5mu\overline{\mkern-1.5mu#1\mkern-1.5mu}\mkern 1.5mu}

\makeatletter
\renewcommand*{\thesection}{\arabic{section}}
\renewcommand*{\thesubsection}{\thesection.\arabic{subsection}}
\renewcommand*{\p@subsection}{}
\renewcommand*{\thesubsubsection}{\thesubsection.\arabic{subsubsection}}
\renewcommand*{\p@subsubsection}{}
\makeatother

\usepackage{titlesec}
\titleformat{\subsection}[runin]
  {\normalfont\normalsize\bfseries}{\thesubsection.}{3pt}{}[.]
\titleformat{\subsubsection}[runin]
  {\normalfont\normalsize\itshape}{\thesubsubsection.}{3pt}{}[\mbox{~}]
\input{macros.tex}

\title{BiD Codes: Algebraic Codes from $3 \times 3$ Kernel}
\author{Anirudh Dash, K. R. Nandakishore, Lakshmi Prasad Natarajan, Prasad Krishnan%
\thanks{This work was conducted when Dash and Nandakishore were with the Department of Electrical Engineering, Indian Institute of Technology Hyderabad, Sangareddy 502248, India (email: anirudh.dash@yahoo.com and nandakishore1331@gmail.com). Natarajan is with the Department of Electrical Engineering, Indian Institute of Technology Hyderabad, Sangareddy 502248, India (email: lakshminatarajan@iith.ac.in). Krishnan is with the Signal Processing and Communications Research Center, International Institute of Information Technology, Hyderabad 500032, India (email: prasad.krishnan@iiit.ac.in).}%
\thanks{This work was supported by the Qualcomm 6G University Research India Program and by ANRF via grant CRG/2023/08696.}% <-this % stops a space
}

\begin{document}

\maketitle

\begin{abstract}
We introduce Berman-intersection-dual Berman (BiD) codes. These are abelian codes of length $3^m$ that can be constructed using Kronecker products of a $3 \times 3$ kernel matrix. BiD codes offer minimum distance close to that of Reed-Muller (RM) codes at practical blocklengths, and larger distance than RM codes asymptotically in the blocklength. Simulations of BiD codes of length $3^5=243$ in the erasure and Gaussian channels show that their block error rates under maximum-likelihood decoding are similar to, and sometimes better, than RM, RM-Polar, and CRC-aided Polar codes.
\end{abstract}

% \begin{IEEEkeywords}
% decoding, minimum weight codewords, product codes, Reed-Muller codes, weight distribution.
% \end{IEEEkeywords}

% % % % % 
% include this in arXiv
% \clearpage
% \setcounter{tocdepth}{2} % do not show subsubsections
% \tableofcontents
% \clearpage

% % % % %

\section{Introduction}

The generator matrices of Reed-Muller (RM) codes~\cite{Muller_IRE_54,Reed_IRE_54,ASY_IT_20} and Ar{\i}kan's Polar codes~\cite{Arikan_IT_2009} are constructed from the same $2 \times 2$ kernel matrix $\p{A}_2 = [1~0;~1~1]$, albeit with different construction rules---RM codes use the Hamming weight of the rows of $\p{A}_2^{\otimes m}$ for construction, while Polar codes use the quality of the bit-channels under successive cancellation (SC) decoding.
While both these codes achieve the capacity of binary-input memoryless symmetric output channels~\cite{AbS_FOCS_2023,ReP_IT_2024,Arikan_IT_2009}, RM codes have a deterministic algebraic construction, a larger minimum distance, and offer lower block error rates (BLER)~\cite{MHU_TCOM_2014} under maximum-likelihood (ML) decoding. 

The construction of RM codes is specific to the $2 \times 2$ kernel; in contrast, the polarization phenomenon is more general and can be induced using kernels of arbitrary sizes~\cite{KSU_IT_10}. The choice of the kernel determines the rate of polarization (effectively, the rate of decay of BLER with respect to the blocklength under SC decoding). 
While the rate of polarization
% \footnote{The rate of polarization tends to $1$ for very large kernel sizes.} 
of $\p{A}_2$ is $0.5$, the best rate of polarization among $3 \times 3$ kernels is close to $0.42$.
Hence, as far as Polar codes are concerned, $3 \times 3$ kernels are less appealing than $\p{A}_2$. 
However, this doesn't preclude the existence of codes based on $3 \times 3$ kernels with  BLER and minimum distance comparable to, or even better than, RM codes. 
This paper is an investigation in this direction. 

We study a family of abelian codes (of length $3^m$) which were designed in~\cite{NaK_IT_23} as ideals in the group algebra of the group $(\Zb_3^m,+)$ over $\Fb_2$. 
We call these codes BiD (Berman-intersection-dual Berman) codes. 
As indicated by this name, they are the intersection of Berman codes~\cite{Ber_Cybernetics_II_67,BlN_IT_01} and the duals of Berman codes. 
These codes achieve vanishing bit error probability (as \mbox{$m \to \infty$}) in the binary erasure channel at all rates less than the channel capacity~\cite{NaK_IT_23}.

We show that BiD codes can be constructed through a $3 \times 3$ kernel $\p{A}_3$ by choosing rows from $\p{A}_3^{\otimes m}$ based on their Hamming weight (Section~\ref{sec:3x3_kernels}). This new construction exposes a recursive structure in their generator matrix, which we exploit to derive bounds on the minimum distance (Section~\ref{sec:mindist}). 
In particular, we show that for any $R \in (0,1)$, as the length $N \to \infty$ there exist BiD codes with rates converging to $R$ and minimum distance growing at least as fast as $N^{0.543}$. This is faster than the asymptotic growth of distance of constant-rate RM codes, which is $N^{0.5}$. 
At practical lengths, several BiD codes have minimum distances similar to those of RM codes.
% We first present an alternative construction of these codes using $3 \times 3$ kernels (Section~\ref{sec:3x3_kernels}), and then identify lower bounds on their minimum distance (Section~\ref{sec:mindist}).

Decoders based on successive cancellation can be adapted to work with BiD codes because of their kernel-based construction. 
We use the successive cancellation ordered search (SCOS) algorithm~\cite{YuC_TCOM_24} (which offers an essentially  maximum-likelihood decoding performance) in our simulations for the binary-input additive white Gaussian noise (BI-AWGN) channel (Section~\ref{sec:decoding}). Simulations (for length $3^5=243$) of the BI-AWGN and the erasure channels show that the BLER of BiD codes are similar to, and sometimes better than, RM codes (of length $256$), CRC-aided Polar codes and RM-Polar codes~\cite{LST_arXiv_14} under maximum-likelihood decoding (Section~\ref{sec:simulations}).

\emph{Notation:} The symbol $\otimes$ denotes the Kronecker product. For any positive integer $\ell$ 
% let $[\ell] = \{1,\dots,\ell\}$, and 
let $\Zb_{\ell} \triangleq \{0,1,\dots,\ell-1\}$ be the ring of integers modulo $\ell$.
% Let $\Zb_{\ell}=\lset \ell \rset$ be the ring where addition and multiplication are performed modulo $\ell$.
% The binary field is $\Fb_2=\{0,1\}$. 
% The empty set is $\emptyset$.
% For sets $A, B$, $A \setminus B \triangleq \left\{a \in A~:~ a \notin B \right\}$, and $\bar{A}$ is the complement of $A$. 
For a set $A$, $\bar{A}$ is its complement.
Capital bold letters denote matrices,  small bold letters denote row vectors, $(\cdot)^T$ is the transpose operator, and $\wt$ denotes the Hamming weight.
We use $\left[\,\p{a}_1; \, \p{a}_2; \, \cdots ; \, \p{a}_\ell\,\right]$ to denote the matrix with rows $\p{a}_1,\dots,\p{a}_{\ell}$.
% All vectors are row vectors, unless otherwise stated. 
% The notation $(.)^T$ denotes the transpose operator.
% The notation $\p 0$ denotes a zero-vector of appropriate size. 
% We denote the identity matrix of size $n$ by ${\p I}_n$. 
% For two vectors ${\p a},{\p b}$, their concatenation is denoted by $({\p a}|{\p b}).$  
The dimension, minimum distance and dual code of a linear code (subspace) $\mathcal{C}$ are $\dimenof(\mathcal{C})$, $\dmin(\mathcal{C})$ and $\mathcal{C}^\perp$, respectively.
% The dimension of a linear code (subspace) $\mathcal{C}$ is denoted by $\dimenof(\mathcal{C})$, its minimum distance by $\dmin(\mathcal{C})$ and its dual code by $\mathcal{C}^\perp$. 
% The Hamming weight of a vector ${\p a}$ is denoted by $\wt({\p a})$. 
The support of a vector $\p{a}$ is $\supp(\p{a})$.
% The minimum distance of a code ${\Cs}$ is denoted by $d_{\min}({\Cs}).$ 
We use $\spanof$ to denote the span of a collection of vectors, and 
$\rowspaceof$ to denote the span of the rows of a matrix.
% The binomial coefficient $\binom{n}{k}$ is assumed to be $0$ if $k>n$ or if $k<0$. 
Finally, $\indicator$ denotes the indicator function.

\section{Algebraic Codes from $3 \times 3$ Kernel} \label{sec:3x3_kernels}

Our first task is to review the spectral-domain construction of abelian codes of length $3^m$ from~\cite[Sec.~IV]{NaK_IT_23} and show that these codes can be constructed in a simpler way---by choosing rows with specific Hamming weights from $\p{A}_3^{\otimes m}$, where $\p{A}_3$ is a binary $3 \times 3$ kernel.

\subsection{Spectral Description of Abelian Codes} \label{subsec:spectral_description_abelian_codes}

% \subsubsection{Spectral Description}

Index the coordinates of binary vectors $\p{a} \in \Fb_2^{3^m}$ using $m$-tuples $\p{i} =(i_1,\dots,i_m) \in \Zb_3^m$, i.e., $\p{a} = \left( a_{\p{i}} : \p{i} \in \mathbb{Z}_3^m \right)$ with $a_{\p{i}} \in \Fb_2$. 
The $m$-dimensional discrete Fourier transform (DFT) of $\p{a}$ yields a length-$3^m$ spectral-domain vector over 
the field $\Fb_4$, 
$\p{\hat{a}} = (\hat{a}_{\p{j}} : \p{j} \in \mathbb{Z}_3^m)$ with $\hat{a}_{\p{j}} \in \Fb_4$. Explicitly, the DFT is given by 
% \begin{equation*}
$\hat{a}_{\p{j}} = \sum_{\p{i} \in \mathbb{Z}_3^m} a_{\p{i}} \, \alpha^{ \p{i} \cdot \p{j} }$,
% \end{equation*}
where $\alpha$ is the primitive element of $\Fb_4$ and $\p{i} \cdot \p{j} = \sum_{\ell=1}^{m} i_\ell j_\ell$ is the inner product of $\p{i}$ and $\p{j}$ over $\Zb_3$~\cite{RaS_IT_92}. The codes of interest to us arise from Remark~36 in~\cite{NaK_IT_23}, and 
are specified by an integer $m$ (that determines the length $3^m$) and a \emph{frequency weight set} $\Wc \subseteq \{0,1\dots,m\}$. The abelian code (denoted by $\abelian$) with parameters $(m,\Wc)$ is
\begin{equation*}
\abelian(m,\Wc) \triangleq \left\{ \p{a} \in \Fb_2^{3^m} \!\! : \! \hat{a}_{\p{j}} = 0 \text{ for all } \p{j} \text{ with } \wt(\p{j}) \notin \Wc \right\} \! .
\end{equation*}
The code corresponding to $\Wc=\emptyset$ is the trivial code $\{\p{0}\}$.

\subsubsection*{Basic Code Properties.} 
If $\Wc' \subset \Wc$ then $\abelian(m,\Wc') \subset \abelian(m,\Wc)$, and 
for any pair of sets $\Wc,\Wc' \subseteq \{0,\dots,m\}$, we have $\abelian(m,\Wc) \cap \abelian(m,\Wc') = \abelian(m,\Wc \cap \Wc')$; these follow immediately from the spectral-domain definition of these codes.
% the observation that $\p{a} \in \abelian(m,\Wc) \cap \abelian(m,\Wc')$ if and only if \mbox{$\hat{a}_{\p{j}}=0$} for all $\p{j}$ with $\wt(\p{j}) \notin \Wc \cap \Wc'$.
Similarly, it is easy to show that $ \abelian(m,\Wc) + \abelian(m,\Wc') = \abelian(m,\Wc \cup \Wc')$.
From these facts it is clear that for any $\Wc$ the subcodes $\abelian(m,\{w\})$, \mbox{$w \in \Wc$}, form a direct-sum decomposition of $\abelian(m,\Wc)$, i.e.,
\begin{equation} \label{eq:direct_sum_decomp}
\abelian(m,\Wc) = \bigoplus_{w \in \Wc} \abelian(m,\{w\}).
\end{equation}
That is, every \mbox{$\p{a} \in \abelian(m,\Wc)$} has a unique representation \mbox{$\p{a}=\sum_{w \in \Wc}\p{a}_w$} where \mbox{$\p{a}_w \in \abelian(m,\{w\})$} for all \mbox{$w \in \Wc$}.
The dual code of $\abelian(m,\Wc)$ is $\abelian(m,\overbar{\Wc})$~\cite[Lemma~27]{NaK_IT_23}.
% these two codes form a complementary pair (\emph{linear complementary dual} codes) since $\Wc \cap \overbar{\Wc} = \emptyset$.

\subsubsection*{Generator Matrix from Inverse DFT.}
Appendix~C of~\cite{NaK_IT_23} gives a construction of a generator matrix that relies on the inverse DFT map. This construction applied to $\abelian(m,\{w\})$ yields the following. Label the columns of the generator matrix using elements of $\Zb_3^m$, and the rows using $\{ \p{j} \in \Zb_3^m : \wt(\p{j})=w \}$. The number of rows is $\binom{m}{w}2^w$. 
% If $w=0$, i.e., when $\p{j}=\p{0}$, all the entries are $1$ (this is the repetition code), and if $w \neq 0$, 
The entry in row $\p{j}$ and column $\p{i}$ is $\indicator\{\p{i}\cdot\p{j} \neq 1 \}$. 
When $w=0$, i.e., $\p{j}=\p{0}$, we get the generator matrix of the repetition code.
When $\p{j} \neq \p{0}$, the support of the $\p{j}^\tth$ row is the union of the $(m-1)$-dimensional subspace $\spanof(\p{j})^\perp$ and exactly one of its two other cosets in $\Zb_3^m$. 
Hence the Hamming weight of this row is $2 \times 3^{m-1}$.
% for every $\p{j} \neq \p{0}$.

The direct-sum decomposition~\eqref{eq:direct_sum_decomp} implies that for any $\Wc \subseteq \{0,1,\dots,m\}$, the vertical concatenation of the generator matrices of $\abelian(m,\{w\})$, \mbox{$w \in \Wc$}, yields a generator matrix for $\abelian(m,\Wc)$. The dimension of $\abelian(m,\Wc)$ is $\sum_{w \in \Wc} \binom{m}{w} 2^w$. 

\subsubsection*{Berman Codes and their Duals.} For $r \in \{-1,0,\dots,m\}$, $\abelian(m,\{r+1,\dots,m\})$ is the $r^\tth$ order Berman code of length $3^m$, and its dual is $\abelian(m,\{0,\dots,r\})$~\cite[Corollary~32]{NaK_IT_23}. 
The minimum distances of these codes are known. 

\begin{theorem}[\cite{Ber_Cybernetics_II_67, BlN_IT_01,NaK_IT_23}] \label{thm:dmin_berman_dual_berman}
The minimum distances of the Berman code $\abelian(m,\{r+1,\dots,m\})$ and its dual $\abelian(m,\{0,\dots,r\})$ are $2^{r+1}$ and $3^{m-r}$, respectively.
\end{theorem}

% they are $2^{r+1}$ and $3^{m-r}$, respectively; see~\cite{NaK_IT_23, Ber_Cybernetics_II_67, BlN_IT_01}.
% the order $r$ Berman code and its dual have minimum distances $2^{r+1}$ and $3^{m-r}$, respectively~\cite{NaK_IT_23}. 

\subsubsection*{Berman-intersection-Dual Berman Codes.}

The principal objects of the current work are the codes obtained as the intersection of a Berman code (of order $r_1-1$) and a dual Berman code (of order $r_2$), where $r_1 \leq r_2$; if $r_1 > r_2$ the intersection is the trivial code $\{\p{0}\}$. 
A binary vector $\p{a}$ lies in the intersection if and only if its spectral components $\hat{a}_{\p{j}}=0$ for all $\p{j}$ with $\wt(\p{j}) \notin \{r_1,r_1+1,\dots,r_2\}$.
This is precisely the code $\abelian(m,\{r_1,\dots,r_2\})$. 

\begin{definition}
The Berman-intersection-dual Berman (BiD) code of length $3^m$ and order parameters $r_1,r_2$ with $0 \leq r_1 \leq r_2 \leq m$ is 
% \begin{equation*}
$\bid(m,r_1,r_2) \triangleq \abelian(m,\{r_1,r_1+1,\dots,r_2\})$.
% \end{equation*}
\end{definition}

% An abelian code is a BiD code when its weight set consists of contiguous integers.
BiD codes have the below direct-sum structure  
\begin{equation*}
\textstyle
\bid(m,r_1,r_2) \! = \bigoplus_{w=r_1}^{r_2} \!\! \bid(m,w,w) 
\! = \bigoplus_{w=r_1}^{r_2} \!\! \abelian(m,\{w\}).
 \end{equation*}
Clearly, $\dimenof \left( \bid(m,r_1,r_2) \right) = \sum_{w=r_1}^{r_2}\binom{m}{w}2^w$. 
Note that $\bid(m,r_1,m)$ is the Berman code of order $r_1-1$, and $\bid(m,0,r_2)$ is the dual Berman code of order $r_2$.
The sum of the order-$r_2$ Berman code and the order-$(r_1-1)$ dual Berman code is precisely the dual of $\bid(m,r_1,r_2)$, which is not a BiD code.
% The dual of $\bid(m,r_1,r_2)$ is the sum of the order-$r_2$ Berman code and the order-$(r_1-1)$ dual Berman code, which is not a BiD code. \pk{Slight rewording needed here perhaps. `The sum of the order-$r_2$ Berman code and the order-$(r_1-1)$ dual Berman code is precisely the dual of $\bid(m,r_1,r_2)$, which is not a BiD code'.}

\subsection{Description via $3 \times 3$ Kernel} 
% To work with the abelian codes of length $N=3^m$, we 
For an alternative description of BiD codes
consider the matrix % $\p{A}_N \triangleq \p{A}_3^{\otimes m}$ where
\begin{equation*} \label{eq:A3}
\p{A}_N \triangleq \p{A}_3^{\otimes m}, \text{ where }
\p{A}_3 \triangleq \left[1~1~1;~1~1~0;~1~0~1\right] \in \Fb_2^{3 \times 3}.
% \begin{bmatrix}
% 1 & 1 & 1 \\ 1 & 1 & 0 \\ 1 & 0 & 1 
% \end{bmatrix}.
\end{equation*}
Since $\p{A}_3$ has linearly independent rows so does $\p{A}_N$.
Each of the $3^m$ rows of $\p{A}_N$ can be uniquely written as a  Kronecker product of $m$ vectors where each of these $m$ vectors is a row of $\p{A}_3$. 
Since the Hamming weight of the Kronecker product of vectors is the product of their individual Hamming weights, the Hamming weight of each row of $\p{A}_N$ is $2^{w} \times 3^{m-w}$ for some $w \in \{0,\dots,m\}$. 
Here \mbox{$w$} is the number of times the second or third rows of $\p{A}_3$ appear as components in the $m$-fold Kronecker product. 
Thus the number of rows of $\p{A}_N$ with Hamming weight $2^w \times 3^{m-w}$ is $\binom{m}{w}2^w$. 
For $w \in \{0,\dots,m\}$, define $\p{G}_{m,w}$ to be the submatrix of $\p{A}_N$ consisting of the rows with Hamming weight $2^w \times 3^{m-w}$. 
% This, we will show, is a generator matrix for $\abelian(m,\{w\})$.

We intend to show that $\p{G}_{m,w}$ is a generator matrix for $\abelian(m,\{w\})$.
Observe that the number of rows in $\p{G}_{m,w}$ is equal to the dimension of $\abelian(m,\{w\})$,
and $\p{G}_{m,w}$ has linearly independent rows (since it is a submatrix of $\p{A}_N$). 
Hence, it is sufficient to prove that each row of the inverse-DFT-based generator matrix of $\abelian(m,\{w\})$ (see Section~\ref{subsec:spectral_description_abelian_codes}) lies in $\rowspaceof(\p{G}_{m,w})$.
Towards this we use the below lemma.
% We use induction on $m$ and 
% the following easily verifiable facts 

\begin{lemma} \label{lem:G_mw_recursion}
For any $m \geq 2$ and $1 \leq w \leq m-1$, we have
\begin{align} \label{eq:G_mw_matrix_recursion}
\p{G}_{m,w} = 
\left[ \!
\begin{array}{l}
(1,1,1) \otimes \p{G}_{m-1,w} \\
(1,1,0) \otimes \p{G}_{m-1,w-1} \\
(1,0,1) \otimes \p{G}_{m-1,w-1}
\end{array}
\! \right]
\text{ and}
\end{align}
$\p{G}_{m,0} = [1~1~1]^{\otimes m} = [1~1~1] \otimes \p{G}_{m-1,0}$, and finally, $\p{G}_{m,m}=[1~1~0;~1~0~1]^{\otimes m} = [1~1~0;~1~0~1] \otimes \p{G}_{m-1,m-1}$.
\end{lemma}
\begin{IEEEproof}
This is an easily verifiable recursion. We have included a proof in Appendix~\ref{app:lem:G_mw_recursion} for completeness.
\end{IEEEproof}

\begin{theorem} \label{thm:single_w_gen_matrix}
For any \mbox{$0 \leq w \leq m$}, the submatrix of $\p{A}_3^{\otimes m}$ consisting of rows with Hamming weight $2^w \times 3^{m-w}$ is a generator matrix for $\abelian(m,\{w\})$.
\end{theorem}
\begin{IEEEproof}
Uses Lemma~\ref{lem:G_mw_recursion} and induction on $m$, 
please see Appendix~\ref{app:thm:single_w_gen_matrix}.
\end{IEEEproof}

Using Theorem~\ref{thm:single_w_gen_matrix} with the direct-sum structure of BiD codes immediately yields the following result.
% $\abelian(m,\Wc)$ yields 
% \begin{corollary} 
% For any $\Wc \subseteq \{0,\dots,m\}$, the submatrix of $\p{A}_3^{\otimes m}$ with Hamming weights equal to $2^w \times 3^{m-w}$ for $w \in \Wc$ is a generator matrix of $\abelian(m,\Wc)$.
% \end{corollary}

% Specializing this to BiD codes we obtain 
\begin{corollary} \label{cor:bid_gen_matrix}
For any $0 \leq r_1 \leq r_2 \leq m$, the rows of $\p{A}_3^{\otimes m}$ with Hamming weights in the range $[2^{r_2} \times 3^{m-r_2}, 2^{r_1} \times 3^{m-r_1}]$ generate $\bid(m,r_1,r_2)$.
\end{corollary}

An identical set of results holds if we permute the rows of $\p{A}_3$ or replace any one of the weight-$2$ rows of $\p{A}_3$ with $(0,1,1)$. One such alternative kernel for BiD codes is $\p{A}'_3 \triangleq [1~1~0;~1~0~1;~1~1~1]$, which is used in our simulations.
% \begin{equation*} \label{eq:A3_dash}
% \p{A}_3' \triangleq 
% \begin{bmatrix}
% 1 & 1 & 0 \\
% 1 & 0 & 1 \\
% 1 & 1 & 1
% \end{bmatrix}.
% \end{equation*}

\section{Bounds on the Minimum Distance of BiD Codes} \label{sec:mindist}

% Our next objective is to identify bounds on the minimum distance of BiD codes. Our technique makes use of the recursion~\eqref{eq:G_mw_matrix_recursion} and the knowledge of the exact minimum distance of Berman and dual Berman codes.
We now derive bounds on $\dmin\left( \bid(m,r_1,r_2) \right)$ using recursion.
We do not have results on the minimum distance of 
$\abelian(m,\Wc)$ for a general choice of $\Wc \subseteq \{0,\dots,m\}$. 
% we do not know which choices of $\Wc \subseteq \{0,\dots,m\}$ provide the best possible trade-off between the code rate and the minimum distance. 
However, we could do significantly worse than BiD codes if $\Wc$ is not chosen carefully; see Appendix~\ref{app:lem:odd_even_weight_set_dmin} for a family of abelian codes with much smaller minimum distance than BiD codes.

% \subsection{Recursive Bounds on the Minimum Distance} 

\begin{figure}[!t]
\centering
\includegraphics[width=\linewidth]{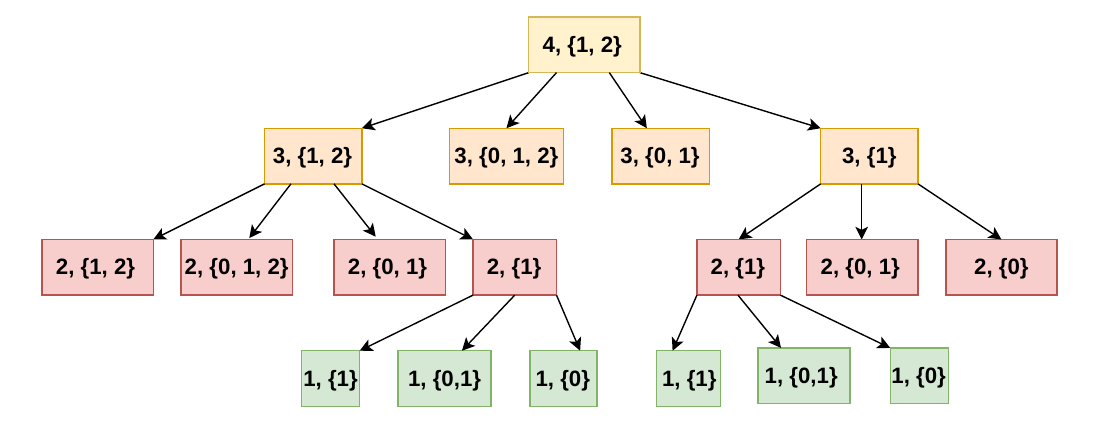}
\caption{The recursion tree for bounding the minimum distance of $\bid(4,1,2)=\abelian(4,\{1,2\})$ via Theorem~\ref{thm:dmin_recursions}. The nodes represent the $(m,\Wc)$ parameters of the BiD codes involved in the recursion.}
\vspace{-5mm}
% \caption{Recursive lower bound calculation tree: Each node indicates the weight set corresponding to the code under consideration. The minimum distance is the minimum over all minimum distances of nodes at the same height. \lp{Update tree to that of $(4,\{1,2\}$}}
\label{fig:min_dist_rec_tree}
\end{figure}

The key idea 
% in deriving a lower bound on the minimum distance of an arbitrary BiD code $\bid(m,r_1,r_2)$ 
lies in the recursive formulation of the generator matrix of $\bid(m,r_1,r_2)$, whose frequency weight set is $\Wc=\{r_1,\dots,r_2\}$. 
% We will denote this generator matrix as $\p{G}_{m,\Wc}$ where the weight set $\Wc=\{r_1,\dots,r_2\}$.
% (call this $\p{G}_{m,\Wc}$, where $\Wc= \left\{ r_1, r_1+1, \cdots r_2 \right\}$ represents the weight set). 
Given a codeword $\p{\rho} \in \bid(m,r_1,r_2) = \abelian(m,\Wc)$, we split it into three truncated subvectors of length $3^{m-1}$ each, i.e., $\p{\rho} = (\p{\rho}_0,\p{\rho}_1,\p{\rho}_2)$. % , where each $\p{\rho}_i$ is of length $3^{m-1}$. 
% In our formulation we split a codeword $\p{\rho} \in \bid(m,r_1,r_2)=\abelian(m,\Wc)$ into three subvectors of length $3^{m-1}$ each \comm{can we add something along the lines of: "Let $\rho$ be a codeword of length $3^m$. Then, "} $\p{\rho} = (\p{\rho}_0,\p{\rho}_1,\p{\rho}_2)$, \comm{each truncated codeword $\p{\rho}_i$ can be expressed in terms of ...}and 
We express each $\p{\rho}_i$ using codewords of BiD codes of length $3^{m-1}$ whose frequency weight sets are related to $\Wc$.
% 
% This formulation allows us to express the codewords of $\bid(m,r_1,r_2)=\abelian(m,\Wc)$ using an appropriate concatenation of codewords of BiD codes of length $3^{m-1}$ whose weight sets are related to $\Wc$. 
An exhaustive list of cases follows (based on the number of non-zero vectors among $\p{\rho}_0,\p{\rho}_1,\p{\rho}_2$) using which we determine upper and lower bounds on the minimum distance of $\bid(m,r_1,r_2)$ by using the minimum distances of the codes of smaller length, effectively creating a recursion tree (see Fig.~\ref{fig:min_dist_rec_tree}). Traversing the recursion along this tree of BiD codes gives us bounds on the minimum distance of all the codes present in the tree.  
% The values obtained for some of these codes are precisely the minimum distance of the codewords, while in others, it is merely a bound. We will later specify the conditions under which such bounds are tight.

We now introduce some notation. 
We will denote the generator matrix of $\abelian(m,\Wc)$ as $\p{G}_{m,\Wc}$. 
When $\Wc=\{w\}$, with a mild abuse of notation, we use $\p{G}_{m,w}$ to denote the generator matrix.
We will assume that the parameters of $\bid(m,r_1,r_2)$ satisfy \mbox{$0 < r_1 \leq r_2 < m$} since the minimum distances for the cases $r_1=0$ (dual Berman) and $r_2=m$ (Berman) are known (Theorem~\ref{thm:dmin_berman_dual_berman}). 
For any given $m$, any weight set $\Wc=\{r_1,\dots,r_2\} \subseteq \{1,2,\dots,m-1\}$, and for appropriate choices of non-negative integers $i,j$, we define $\Wc_{-i,-j} \triangleq \{r_1-i,r_1-i+1,\dots,r_2-j\}$. We will use $\Wc_{-i,-j}$ as the frequency weight set of a BiD code of length $3^{m-1}$. 

From the direct-sum structure~\eqref{eq:direct_sum_decomp} observe that $\p{G}_{m,\Wc}$ is the vertical concatenation of $\p{G}_{m,w}$, $w \in \Wc$. Since $r_1 > 0$ and $r_2 < m$, we observe that $w \in \{1,\dots,m-1\}$ for all $w \in \Wc$, and hence, the recursion~\eqref{eq:G_mw_matrix_recursion} applies to each $\p{G}_{m,w}$. 
Now applying the direct-sum structure~\eqref{eq:direct_sum_decomp} to the length-$3^{m-1}$ codes arising from the recursion~\eqref{eq:G_mw_matrix_recursion} we immediately obtain 
\begin{align*} %\label{eq:G_mW_recursion}
\p{G}_{m,\Wc} = 
\left[ 
\begin{array}{l}
(1,1,1) \otimes \p{G}_{m-1,\Wc} \\
(1,1,0) \otimes \p{G}_{m-1,\Wc_{-1,-1}} \\
(1,0,1) \otimes \p{G}_{m-1,\Wc_{-1,-1}}  
\end{array}
\right]. 
\end{align*}
This implies that for every $\p{\rho} \in \abelian(m,\Wc)$ there exist unique $\p{a} \in \abelian(m-1,\Wc)$ and $\p{a}',\p{a}'' \in \abelian(m-1,\Wc_{-1,-1})$ such that 
\begin{equation} \label{eq:rho_kronecker_decomp}
\p{\rho} = (1,1,1) \otimes \p{a} + (1,1,0) \otimes \p{a}' + (1,0,1) \otimes \p{a}''.   
\end{equation}
We now express $\p{a},\p{a}',\p{a}''$ using the direct-sum decomposition~\eqref{eq:direct_sum_decomp} as follows 
\begin{equation} \label{eq:a_direct_sum_decomp}
\textstyle 
\p{a} = \sum_{k=r_1}^{r_2} \p{a}_k,
\p{a}' = \! \sum_{k=r_1-1}^{r_2-1} \! \p{a}_k' \text{ and }
\p{a}'' = \! \sum_{k=r_1-1}^{r_2-1} \! \p{a}_k'',
\end{equation}
where $\p{a}_w,\p{a}'_w,\p{a}_w'' \in \abelian(m-1,\{w\})$ for each integer $w$.
If $\p{\rho}$ is split into subvectors as $(\p{\rho}_0,\p{\rho}_1,\p{\rho}_2)$, from~\eqref{eq:rho_kronecker_decomp} and~\eqref{eq:a_direct_sum_decomp},
\begin{align} \label{eq:rho_012}
\begin{split}
\textstyle 
\p{\rho}_0&=\sum_{k=r_1}^{r_2-1} (\p{a}_k+\p{a}_k'+\p{a}_k'') + \p{a}_{r_2} + \p{a}_{r_1-1}' + {\p{a}}_{r_1-1}'', \\
\p{\rho}_1&=\sum_{k=r_1}^{r_2-1} (\p{a}_k+\p{a}_k') + \p{a}_{r_2} + \p{a}_{r_1-1}', \text{ and} \\ 
\p{\rho}_2&=\sum_{k=r_1}^{r_2-1} (\p{a}_k+\p{a}_k'') + \p{a}_{r_2} + {\p{a}}_{r_1-1}''. 
\end{split}
\end{align} 
We consider an exhaustive list of cases (based on the number of non-zero vectors among $\p{\rho}_0,\p{\rho}_1,\p{\rho}_2$), use the sum-structure in~\eqref{eq:rho_012} and the direct-sum decomposition~\eqref{eq:direct_sum_decomp} to bound the distance of $\bid(m,r_1,r_2)$. 
In each of these cases, we obtain bounds on the smallest possible non-zero weight of $\p{\rho}$ using bounds on the weights of $\p{\rho}_i$ by using the fact that each $\p{\rho}_i$ is a codeword of a BiD code of length $3^{m-1}$.

Let $d_{m}(\Wc)$ denote the minimum distance of $\abelian(m,\Wc)$.

\begin{theorem}[Recursive bounds on minimum distance]
\label{thm:dmin_recursions}
Let \mbox{$\Wc = \{r_1,\dots,r_2\}$} with \mbox{$r_1> 0$} and \mbox{$r_2<m$}. Define 
% \emph{(i)}
\mbox{$D_2 = d_{m-1}(\Wc_{0,-1})$} if \mbox{$\lvert \Wc \rvert > 1$} and \mbox{$D_2=+\infty$} otherwise, 
% \emph{(ii)}~
\mbox{$D_3 = 2d_{m-1}(\Wc_{-1,-1})$}, 
% \emph{(iii)}~
\mbox{$D_{4a} = 3d_{m-1}(\Wc_{-1,0})$}, 
$D'_{4} = 3d_{m-1}(\Wc)$, 
% \begin{align*}
% \text{(iv)~} 
$D_{4b} = \min\{D'_4, \, d_{m-1}(\Wc_{-1,-1})+d_{m-1}(\Wc_{-1,0})\}$, and 
% \end{align*}
$D_{4} = \max\{D_{4a},D_{4b}\}$.
% \begin{enumerate}
%     \item $D_2 = d_{m-1}(\Wc_{0,-1}) $ if $\lvert \Wc \rvert > 1$
%     \item $D_3 = 2d_{m-1}(\Wc_{-1,-1})$
%     \item $D_{4a} = 3d_{m-1}(\Wc_{-1,0})$
%     \item $D_{4b}=d_{m-1}(\Wc_{-1,-1})+d_{m-1}(\Wc_{-1,0})$
%     \item $D_{4c} = 3d_{m-1}(\Wc_{0,0})$
%     \item $D_4=\max \{D_{4a}, \min \{D_{4b}, D_{4c} \} \}$
%     \end{enumerate}
Then, 
\begin{align} \label{eq:dmin_recursion_bounds}
% \begin{cases}
%   \min \{D_2, D_3, D_4 \} & \text{if } \lvert \Wc \rvert > 1 \\
%   \min \{D_3, D_4 \} & \text{if } \lvert \Wc \rvert = 1,
% \end{cases} \text{ and }
\min\{D_2,D_3,D_4\} \leq d_m(\Wc) \leq \min\{D_2,D_3,D'_4\}.
\end{align}
\end{theorem}
\begin{IEEEproof}
Please see Appendix~\ref{app:thm:dmin_recursions}.
\end{IEEEproof}

Theorem~\ref{thm:dmin_recursions} can be used recursively to numerically compute bounds on $\dmin(\bid(m,r_1,r_2))$. If $r_1=0$ (dual Berman code) or $r_2=m$ (Berman code), the exact minimum distance is known (Theorem~\ref{thm:dmin_berman_dual_berman}). Otherwise, we can use~\eqref{eq:dmin_recursion_bounds} to recurse to BiD codes of length $3^{m-1}$ with frequency weight sets $\Wc,\Wc_{0,-1},\Wc_{-1,0}$ and $\Wc_{-1,-1}$. 
This naturally leads to a recursion tree, see Fig.~\ref{fig:min_dist_rec_tree}.
The termination cases of the recursion are Berman and dual Berman codes. 
Appendix~\ref{app:list_of_bid_codes} presents these numerically computed bounds on $\dmin$ for all BiD codes of lengths $9$ till $729$.
The computed upper and lower bounds are equal (and hence, yield the exact minimum distance) for all codes of length up to $3^5$ except the following four choices of $(m,r_1,r_2)$: $(4,2,2)$, $(5,2,2)$, $(5,2,3)$, $(5,3,3)$. 

As a corollary to Theorem~\ref{thm:dmin_recursions}, we provide two closed form expressions for lower bounding $\dmin(\bid(m,r_1,r_2))$. The first is a generalization of the minimum distance expression for dual Berman codes, and the second extends the expression for Berman codes to BiD codes. 
% max(4.0^(r1) * 3.0^(m - r1 - r2), 3.0^(m - r2) * 2.0^(r1 + r2 - m))
\begin{theorem} \label{thm:bid_dmin_thoeretical_lower_bound}
The minimum distance of $\bid(m,r_1,r_2)$ is at least 
% lower bounded by 
% \begin{equation*}
$\lceil \max\left\{4^{r_1} \times 3^{m-r_1-r_2}, \, 3^{m-r_2} \times 2^{r_1 + r_2 - m} \right\} \rceil$.
% \end{equation*}
\end{theorem}
\begin{IEEEproof}
The proof uses induction in a straightforward manner. Please see Appendix~\ref{app:thm:bid_dmin_theoretical_lower_bound}.
\end{IEEEproof}

Whenever \mbox{$r_1 \log_2(3/2) + r_2 > m$}, the second bound in Theorem~\ref{thm:bid_dmin_thoeretical_lower_bound} is larger (and hence tighter), and the first expression is a better bound otherwise. 
% \comm{should we specify $\leq m$ here?}
The closed form bound of Theorem~\ref{thm:bid_dmin_thoeretical_lower_bound} is equal to the numerically computed lower bound (that uses Theorem~\ref{thm:dmin_recursions} recursively) for BiD codes of length up to $3^9$ except $\bid(8,5,5)$ and $\bid(9,5,6)$.

Finally we present results for the exact minimum distance for two specific weight sets, $\Wc=\{1\}$ and $\Wc=\{m-1\}$. 

% Finally, we show the calculations for the exact minimum distance of two specific weight sets, $\Wc=\{1\}$ and $\Wc=\{m-1\}$. 
% This corresponds to the codes $\bid(m,1,1)$, $\bid(m,m-1,m-1)$.
% We consider the former case first. 

% Using Theorem \ref{thm:recursive dmin}, 
% \begin{enumerate}
%     \item $D_5=3^{m-1-0}+3^{m-1-1}=4\times 3^{m-2}$
%     \item $D_4:\min \{ D_5, 3d_{m-1}(\Wc_{0,0}) \}$. Even though we do not know $3d_{m-1}(\Wc_{0,0})$, we can keep applying this recursion until the point where we have $3d_{m-(m-2)-1}(\Wc_{0,0})$, which is a Berman code. Thus, this value is $3\times2^1=6$, while $D_5=4$, which is the minimum value.  
%     \item $D_3= \max \{ 3\times 3^{m-1-1}=3^{m-1}, D_4$ \}. This maximum is thus $4\times3^{m-2}$.
%     \item $D_2=2\times 3^{m-1}$
%     \item $D_1:$ invalid case for $\Wc_{0,0}$ such that $\lvert \Wc_{0,0} \rvert = 1$ where $\lvert S \rvert$ denotes the cardinality of set $S.$
% \end{enumerate}

% Thus, the minimum distance is $4\times 3^{m-2}$. This also matches the tightness result described above as here, $r_1=r_2=1$; $r_1+r_2\leq m \ \forall m \geq 2$. 

\begin{theorem} \label{thm:W_equal_to_1}
The minimum non-zero weight and the maximum weight among all codewords of $\bid(m,1,1)$, $m \geq 2$, are $4 \times 3^{m-2}$ and $6 \times 3^{m-2}$, respectively.
\end{theorem}
\begin{IEEEproof}
The proof technique is different from that of Theorem~\ref{thm:dmin_recursions} and relies on the maximum weight codewords in the code. Please see Appendix~\ref{app:thm:W_equal_to_1}.
\end{IEEEproof}

% Let the inductive hypothesis for the minimum distance be $d_{m-1}(\Wc_{0,0}) = 4\times3^{(m-1)-2}$. Similarly, let the inductive hypothesis for the maximum weight codeword in the code be $(\bar{d})_{m-1}(\Wc_{0,0}) = 6\times3^{(m-1)-2}$.

% \lp{several minor typos in this paragraph? also, the induction base case needs to be clarified?}
% Now, note that $\abelian(m,1,1) \subseteq \abelian(m,0,1)$ (which is a Dual Berman code). Thus, $d_{m-1}(\Wc_{0,0}) \geq d_{m-1}(\Wc_{-1,0})=3^{m-2}$. $\p{a}_{m-1,\{1\}} + \p{1} \in \abelian(m,0,1)$. Here, $\p{1}$ is the all ones vector of length $3^{m-1}$. Now, $\wt(\p{a}_{m-1,\{1\}}) = 3^{m-1}-\wt(\p{a}_{m-1),\{1\}} \geq 3{m-2}$. Consequently, $\wt(\p{a}_{m-1,\{1\}} \leq 2\times 3^{m-2}$. $d_{m-1}(\Wc_{0,0})=\min \{3d_{m-1}(\Wc_{0,0}),  2\times3^{m-1}-{\bar{d}}_{m-1}(\Wc_{0,0})\}$. The second term in the expression is at least $4\times 3^{m-2}$. Thus, the minimum is $4\times 3^{m-2}$. An identical approach gives $\bar{d}(\Wc_{0,0}) = 6\times3^{(m-1)-2}.$ 

% Now we consider the weight set $\Wc=\{m-1\}$.

\begin{theorem} \label{thm:W_m_minus_1}
For any $m \geq 3$, the minimum distance of $\bid(m,m-1,m-1)$ is $3 \times 2^{m-2}$.
\end{theorem}
\begin{IEEEproof}
This is a direct application of Theorem~\ref{thm:dmin_recursions} using induction. Please see Appendix~\ref{app:thm:W_m_minus_1}.
\end{IEEEproof}

\subsection*{Asymptotic Growth of Minimum Distance}

% Suppose we want to construct a linear code of length $2^m$ by choosing a collection of rows of $\p{A}_2^{\otimes m}$ as the generator matrix, where $\p{A}_2$ is Ar{\i}kan's $2 \times 2$ kernel. It is well known that choosing the rows according to the RM rule gives the largest possible minimum distance. 
For any $R \in (0,1)$, and for increasing blocklengths $N=2^m$, there exists a sequence of $\RM(m,r)$ codes with rate converging to $R$. For these codes $r/m \to 0.5$; see~\cite{Arikan_IT_2009,KKMPSU_IT_17}. 
Since the minimum distance of $\RM(m,r)$ is $2^{m-r}$, we see that $\log(d_{\min})/\log(N) \to 0.5$ for RM codes. 
This is the best asymptotic growth of $d_{\min}$ for codes constructed from $\p{A}_2^{\otimes m}$ since choosing rows from $\p{A}_2^{\otimes m}$ according to the RM rule gives the largest possible minimum distance~\cite{Korada_thesis}.
% For the $3 \times 3$ kernel $\p{A}_3$, we will show that BiD codes 
% In this work we consider the $3 \times 3$ kernel $\p{A}_3$, and show that BiD codes (whose generator matrices are constructed from $\p{A}_3^{\otimes m}$) have a faster asymptotic growth of $\dmin$. 
We rely on Theorem~\ref{thm:bid_dmin_thoeretical_lower_bound} and a technique similar to~\cite[Remark~24]{KKMPSU_IT_17} to show that BiD codes have a faster asymptotic growth of $\dmin$. The proof of the following result is in Appendix~\ref{app:thm:bid_asymptotic_dmin}.
% to show that BiD codes, that arise from $\p{A}_3^{\otimes m}$, have a faster asymptotic growth.

\begin{theorem} \label{thm:bid_asymptotic_dmin}
For any $R \in (0,1)$ there exists a sequence of BiD codes with $N \to \infty$, rate converging to $R$ and
\begin{equation*}
\liminf_{N \to \infty} \left( \log \left( d_{\min} \right) / \log N \right)\geq \log 6 / \log 27 > 0.543.
\end{equation*}
% and the rate converges to $R$.
\end{theorem}
% \begin{IEEEproof}
% Please see Appendix~\ref{app:thm:bid_asymptotic_dmin}.
% \end{IEEEproof}

\begin{figure}[!t]
    \centering
    \includegraphics[width=\linewidth]{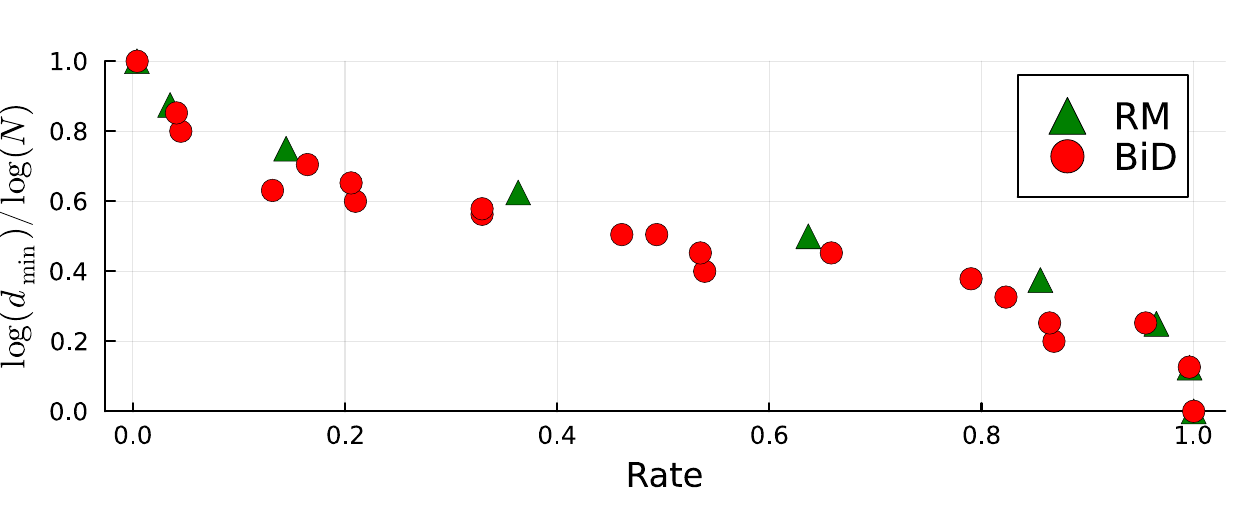}
    \vspace{-3mm}
    \caption{The $\log$-normalized minimum distance of RM codes (length $256$) and BiD codes (length $243$, using Theorem~\ref{thm:bid_dmin_thoeretical_lower_bound}) versus rate.}
    \vspace{-3mm}
    \label{fig:min_dist_scatter_plot}
\end{figure}

Fig.~\ref{fig:min_dist_scatter_plot} compares $\log(\dmin)/\log(N)$ versus rate of all RM codes of length $256$ with all BiD codes of length $243$. 
The exact minimum distance is used for RM codes, while the lower bound from Theorem~\ref{thm:bid_dmin_thoeretical_lower_bound} is used for BiD codes. 
BiD codes provide more options for rate and offer a graceful trade-off of rate versus $d_{\min}$.
While Theorem~\ref{thm:bid_asymptotic_dmin} guarantees a better trade-off between $d_{\min}$ and rate for BiD codes for sufficiently large $N$, BiD codes come close to RM codes even at practical lengths. 
Note that this comparison is made using only a lower bound on $d_{\min}$ for BiD codes, which might be loose.

\section{Decoding} \label{sec:decoding}

% As a consequence of Corollary~\ref{cor:bid_gen_matrix}, 
BiD codes can be viewed in the framework of Polar codes with $3 \times 3$ kernels. % such as $\p{A}_3$~\eqref{eq:A3} and $\p{A}'_3=[1~1~0;~1~0~1;~1~1~1]$.
% and 
The decoders designed for Polar codes can be adapted for BiD codes by the appropriate choice of the locations of frozen bits. 
For instance, if the kernel $\p{A}'_3$ is used, the Polar encoding transform of length $N=3^m$ is $\p{G}'_N \triangleq \p{B}_N \, {\p{A}'_3}^{\otimes m} = {\p{A}'_3}^{\otimes m} \p{B}_N $, where $\p{B}_N$ is the involutary permutation matrix that exchanges the indices $1+\sum_{\ell=1}^{m}i_{\ell} 3^{\ell-1}$ and $1+\sum_{\ell=1}^{m} i_{\ell} 3^{m-\ell}$ for every $(i_1,\dots,i_m) \in \Zb_3^m$. 
The $i^\tth$ bit of $\bid(m,r_1,r_2)$ is frozen if the Hamming weight of the $i^\tth$ row of $\p{G}'_N$ lies outside the range $[2^{r_2} \times 3^{m-r_2}, 2^{r_1} \times 3^{m-r_1}]$, see Corollary~\ref{cor:bid_gen_matrix}.
% As in the $2 \times 2$ case~\cite{Arikan_IT_2009,Pfister_Notes_Polar}, it is straightforward to show that $\p{B}_N {\p{A}'}_3^{\otimes m} = {\p{A}'}_3^{\otimes m}\p{B}_N$.
In contrast, the frozen bits of the Polar code are chosen based on the quality of the bit-channels under successive-cancellation (SC) decoding. 
% the frozen bits of BiD codes are chosen based on the Hamming weight of the rows of $\p{G}'_N$: the $i^\tth$ bit of $\bid(m,r_1,r_2)$ is frozen if the Hamming weight of the $i^\tth$ row of $\p{G}'_N$ lies outside the range $[2^{r_2} \times 3^{m-r_2}, 2^{r_1} \times 3^{m-r_1}]$; see Corollary~\ref{cor:bid_gen_matrix}. 
This is similar to the relation between Polar codes of length $2^m$ and RM codes.

In our simulations for the BI-AWGN channel, we adapt the successive cancellation ordered search (SCOS) decoder~\cite{YuC_TCOM_24,YuC_github} to decode BiD codes; please see Appendix~\ref{app:decoding} for details. 
The complexity of the SCOS decoder is determined by two parameters, denoted as $\lambda_{\max}$ and $\eta$ in~\cite{YuC_TCOM_24}. 
We set $\lambda_{\max}$ and $\eta$ to sufficiently large values so that the block error rate (BLER) of the SCOS decoder is close to that of the ML decoder. 
% \comm{do parameters need to be specified here?}
We verified this by empirically estimating a lower bound on the BLER of the ML decoder, as done in~\cite{YuC_TCOM_24,TaV_IT_15}.  

\begin{figure}[!t]
    \centering
    \includegraphics[width=\linewidth]{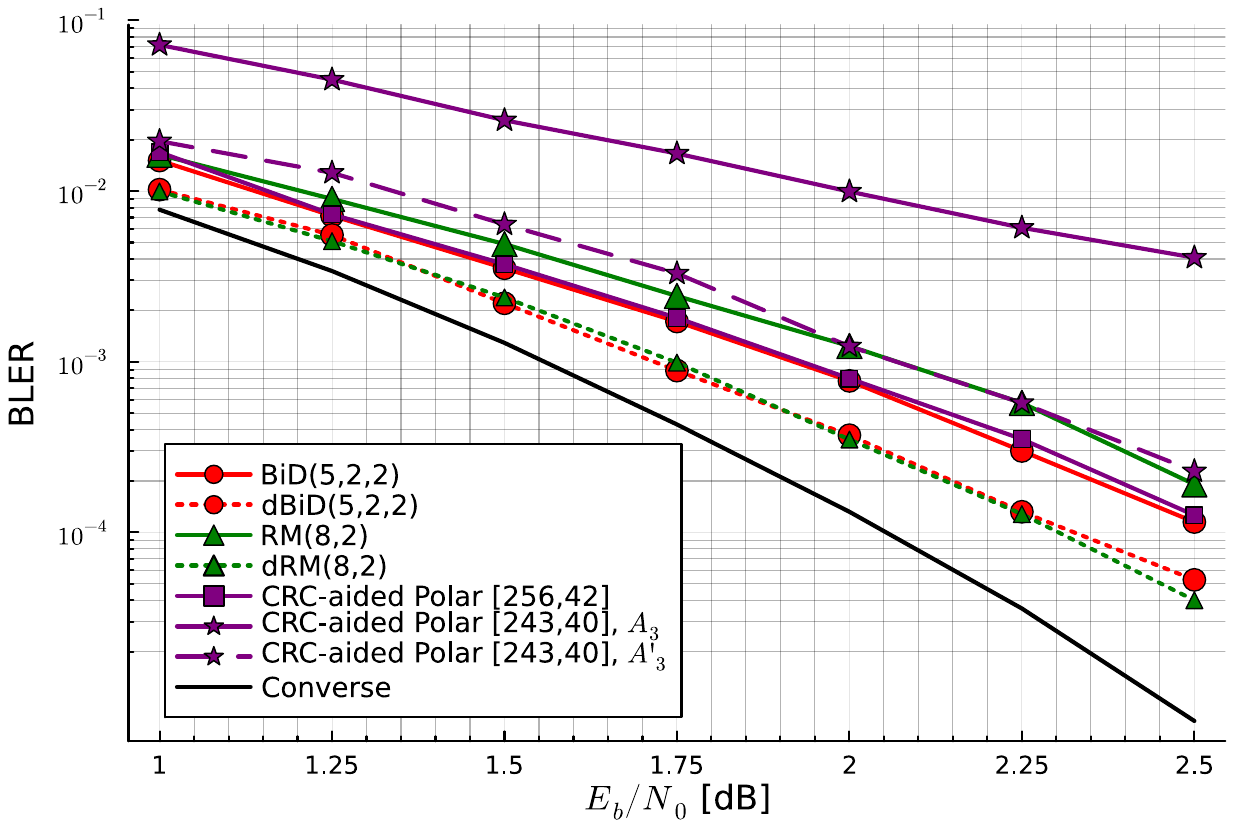}
    \vspace{-5mm}
    \caption{BLER comparison in the BI-AWGN channel.}
    \vspace{-5mm}
    \label{fig:awgn}
\end{figure}

% \begin{figure}[!t]
%     \centering
%     % \begin{minipage}{0.475\linewidth}   
%     \includegraphics[width=\linewidth]
%     {Images/sims/scatter_plot_no_df_labelled.pdf}
%     \caption{Normalized rates versus code rate in the BEC for ${\rm BLER}=10^{-3}$. The $(m,r_1,r_2)$ parameters of some BiD codes are shown.}
%     \label{fig:Rnorm_without_dfcons}
% \end{figure}
% \begin{figure}[!t]
%     % \end{minipage}
%     % \hfill
%     % \begin{minipage}{0.475\linewidth}
%     \includegraphics[width=\linewidth] 
%     {Images/sims/scatter_plot_df_annotated.pdf}
%     \caption{Normalized rates {with} dynamic frozen bits versus code rate at ${\rm BLER}=10^{-3}$ in the BEC.}
%     \label{fig:Rnorm_with_dfcons}
%     % \end{minipage}
% \end{figure}

\begin{figure}[!t] \label{fig:Rnorm_BEC}
    \centering
    % \begin{minipage}{0.475\linewidth}   
    \includegraphics[width=\linewidth]
    {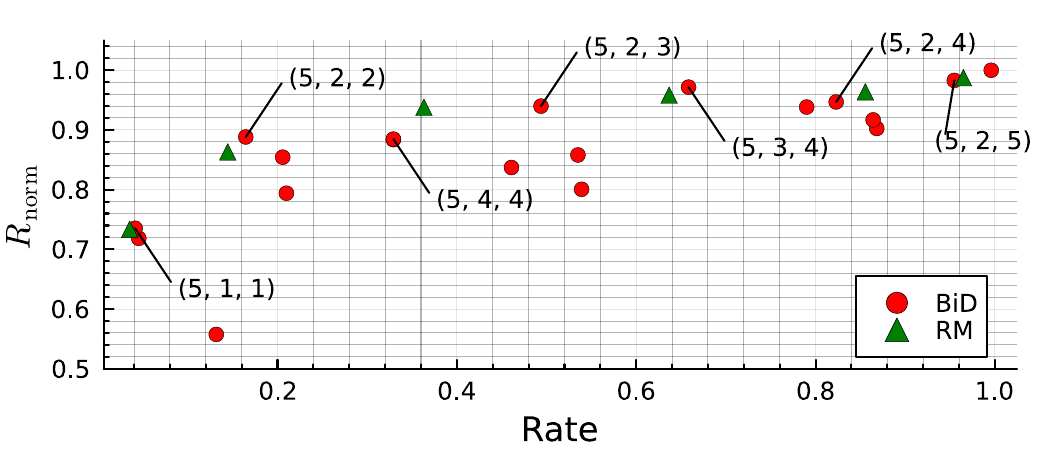} \\
    \vspace{-5mm}
    % \caption{Normalized rates versus code rate in the BEC for ${\rm BLER}=10^{-3}$. The $(m,r_1,r_2)$ parameters of some BiD codes are shown.}
    % \label{fig:Rnorm_without_dfcons}
% \end{figure}
% \begin{figure}[!t]
    % \end{minipage}
    % \hfill
    % \begin{minipage}{0.475\linewidth}
    \includegraphics[width=\linewidth] 
    {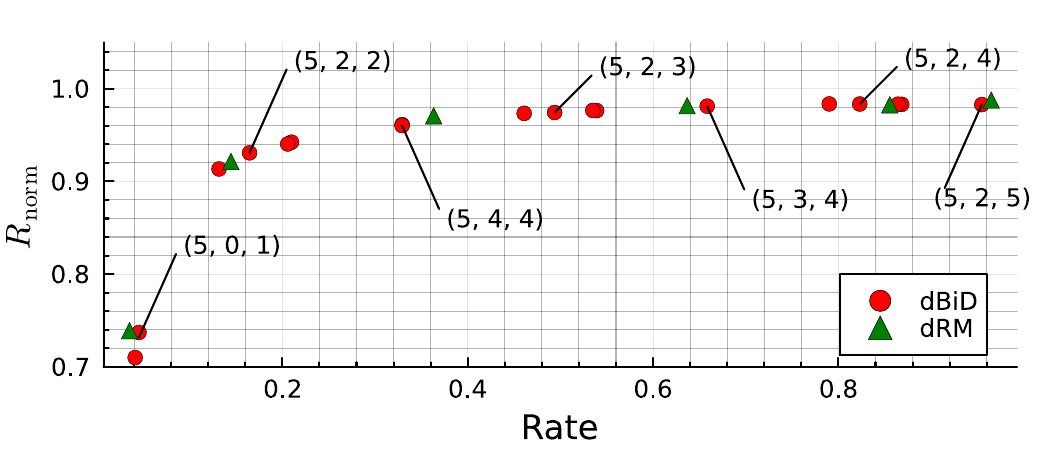}
    \vspace{-7mm}
    \caption{$R_{\rm norm}$ versus code rate in the BEC without (above) and with (below) dynamic freezing. The $(m,r_1,r_2)$ parameters of some BiD codes are shown.}
    % \caption{Normalized rates {with} dynamic frozen bits versus code rate at ${\rm BLER}=10^{-3}$ in the BEC.}
    % \label{fig:Rnorm_with_dfcons}
    % \end{minipage}
    \vspace{-5mm}
\end{figure}

\begin{figure}[!t]
    \centering
        \includegraphics[width=\linewidth]{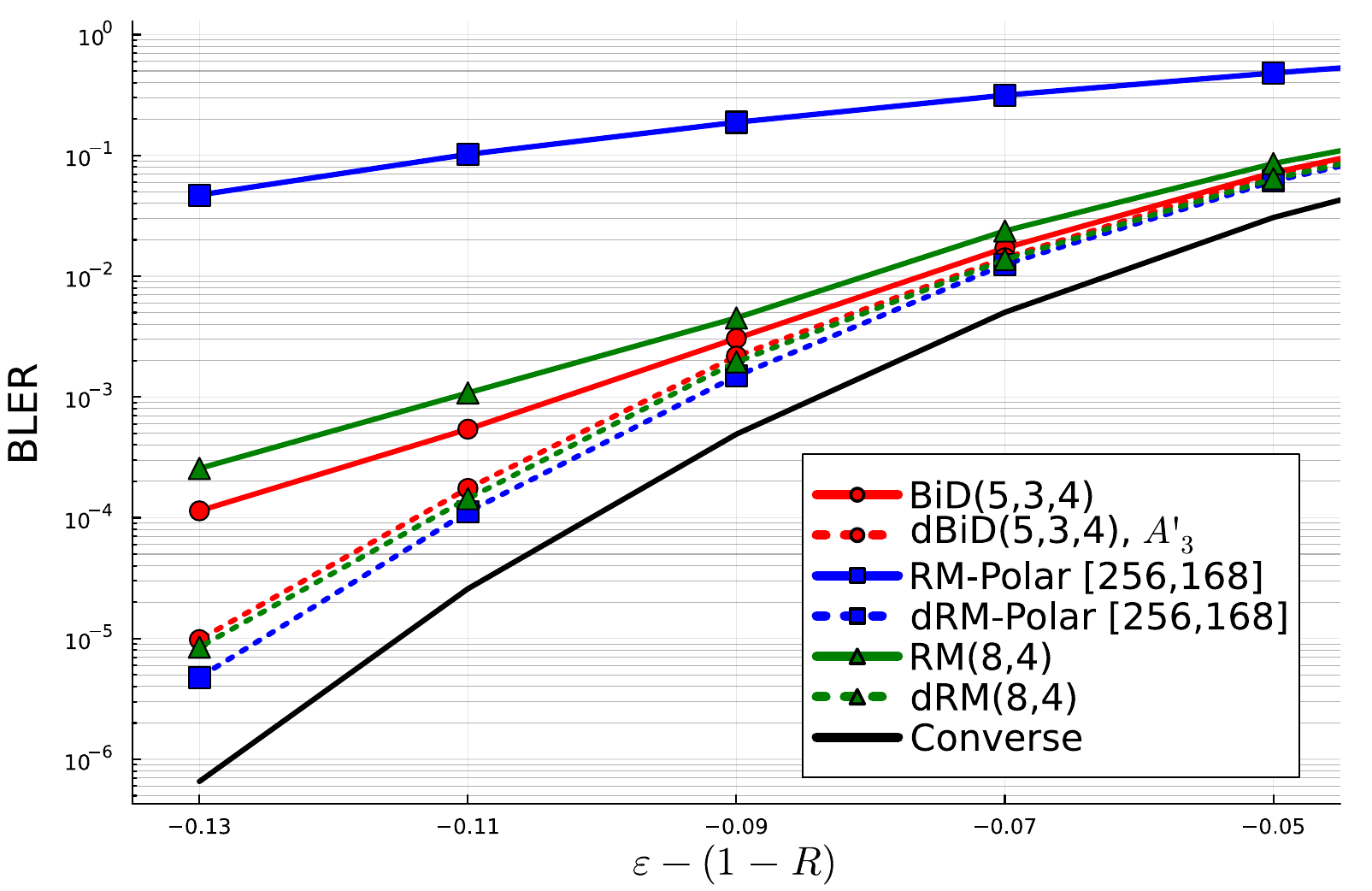}
        \vspace{-5mm}
    \caption{BLER of $\bid(5,3,4)$ and ${\rm d}\bid(5,3,4)$ (kernel $\p{A}'_3$) in BEC.}
    \label{fig:534}
    \vspace{-3mm}
\end{figure}

\section{Simulation Results} \label{sec:simulations}

We present simulation results for the BI-AWGN channel (under SCOS decoding with an essentially-ML performance of BLER) and the erasure channel (under ML decoding)  for BiD codes with length $3^5=243$. We choose this length for the convenience of comparing with codes of length $2^8=256$ (which is close to $243$) obtained from Ar{\i}kan's $2 \times 2$ kernel.
Our objective is to show the ML performance of the new codes without any consideration towards reducing the decoding complexity. 
Unless otherwise mentioned, we use SCOS decoding~\cite{YuC_TCOM_24} for \emph{all the codes} (from both $2 \times 2$ and $3 \times 3$ kernels) in the BI-AWGN channel with large enough $\lambda_{\max}$ and $\eta$ to ensure essentially-ML BLER.

We compare BiD codes with RM codes and RM-Polar codes~\cite{LST_arXiv_14} (both of which are constructed from the $2 \times 2$ kernel) with and without dynamic freezing~\cite{Arikan_PAC_2019,TrM_JSAC_16,YFV_Entropy_21,RBV_TVT_21}.
% codes obtained from the $2 \times 2$ kernel using the RM rule~\cite{MHU_TCOM_2014,VABM_SCVT_2016} (i.e., RM codes) and the RM-Polar rule~\cite{LST_arXiv_14}, both with and without dynamic freezing~\cite{Arikan_PAC_2019,TrM_JSAC_16,YFV_Entropy_21,RBV_TVT_21}. 
We also compare with the CRC-aided Polar codes (kernels $\p{A}_2$, $\p{A}_3$ and $\p{A}'_3$) in the AWGN channel. 
% \lp{The simulation results show that BiD codes are competitive with respect to these codes under ML decoding. With dynamic freezing BiD codes perform close to the finite-length information theoretic lower bounds on BLER.}

% \subsection{Dynamically Frozen Bits}

For the \mbox{$2 \times 2$} kernel, it is known that using dynamic freezing or pre-transformation~\cite{Arikan_PAC_2019,TrM_JSAC_16,YFV_Entropy_21,RBV_TVT_21,LZG_arXiv_19,LZLWYM_ISIT_21,RoY_JSAIT_23} offers dramatic reductions in BLER with performance close to finite-length information theoretic limits on channel coding rate~\cite{PPV_IT_10,Ers_IT_16,spectre_github}. 
With dynamic freezing each frozen bit is set to be the parity of a subset of the bits that are decoded before this bit in the SC decoding order. 
We implement dynamic freezing by using a full-rank upper triangular pre-transformation matrix (denoted as $\p{T}$ in~\cite[Section~VII]{Arikan_PAC_2019}) whose entries above the diagonal are Bernoulli(1/2) distributed; this corresponds to pre-transformation using a time-varying rate-$1$ convolutional code with unbounded constraint length~\cite{YuC_TCOM_24,Arikan_PAC_2019,YFV_Entropy_21}.
We use a leading `d' to denote the dynamically frozen variant of the codes, such as ${\rm d}\RM$, ${\rm d}\bid$.

% For the $2 \times 2$ kernel, dynamic freezing reduces the number of minimum weight codewords without decreasing the minimum distance of the code~\cite{LZG_arXiv_19,LZLWYM_ISIT_21,RoY_JSAIT_23,YFV_Entropy_21}.
% We implement dynamic freezing in our simulations using a random upper triangular matrix $\p{T}$ that transforms the data carrier word before the Polar encoding step, as in~\cite{Arikan_PAC_2019}. All the diagonal elements of $\p{T}$ are set to $1$, all entries above the diagonal are chosen from Bernoulli(1/2) distribution; this corresponds to pre-transformation using a time-varying rate-$1$ convolutional code with unbounded constraint length~\cite{YuC_TCOM_24,Arikan_PAC_2019,YFV_Entropy_21}.
% We apply dynamic freezing to BiD codes too, although there is no guarantee that dynamic freezing does not deteriorate the weight distribution of these codes. 
% We use a leading `d' to denote the dynamically frozen variant of the codes (such as ${\rm d}\RM$ and ${\rm d}\bid$).

\subsection{BI-AWGN Channel}

Fig.~\ref{fig:awgn} shows the BLER versus $E_b/N_0$ curves for 
\begin{enumerate}[label=(\roman*)]
\item $\bid(5,2,2)$ (this is a $[243,40,48]$ code);
\item $\RM(8,2)$ ($[256,37,64]$ code, with RPA-list decoding~\cite{YeA_IT_20} that has near-ML BLER);
% \item the $[256,42,32]$ subcode of $\RM(8,3)$ obtained from the $2 \times 2$ kernel using the RM rule (same rate as $\bid(5,2,2)$);
\item the dynamically frozen variants of these two codes;
\item the $[256,42]$ CRC-aided Polar code with $8$-bit CRC, constructed using the kernel $\p{A}_2$;
\item the $[243,40]$ CRC-aided Polar code with $8$-bit CRC constructed using the kernel $\p{A}_3$; % (this kernel has the largest rate of polarization~\cite{KSU_IT_10} among $3 \times 3$ kernels);
\item same as above, but constructed using the kernel $\p{A}'_3$; % the $[243,40]$ CRC-aided Polar code with $8$-bit CRC constructed using the kernel $\p{A}'_3$;
\item the $\order(N^{-2})$ approximation to the Polyanskiy-Poor-Verd\'{u} metaconverse lower bound on the optimal BLER~\cite[Fig.~6]{Ers_IT_16} (implemented in~\cite{spectre_github}) computed for the rate and length of $\bid(5,2,2)$ and ${\rm d}\bid(5,2,2)$ in BI-AWGN.
\end{enumerate}

All the codes in Fig.~\ref{fig:awgn} have the same rate ($\approx 0.164$) except $\RM(8,2)$ and ${\rm d}\RM(8,2)$ (rate $\approx 0.144$).

% The figure also shows the BLER of the dynamically frozen variants of the first three codes.
% pre-transformed versions of the first three codes obtained via dynamically frozen bits (labelled using a leading `d' in the legend). 
We used the kernel $\p{A}'_3$ for SCOS decoding and dynamic freezing of  $\bid(5,2,2)$ and ${\rm d}\bid(5,2,2)$. 
% use the kernel $\p{A}_3'$ for SCOS decoding and dynamic freezing. 
While Theorem~\ref{thm:bid_dmin_thoeretical_lower_bound} guarantees a minimum distance of at least $48$ for $\bid(5,2,2)$, from our simulations we identified codewords with Hamming weight exactly $48$ in this code. Thus, Theorem~\ref{thm:bid_dmin_thoeretical_lower_bound} is tight for $\bid(5,2,2)$. With dynamic freezing, we observed codewords with weight $27$ in ${\rm d}\bid(5,2,2)$.
% Hence, dynamic freezing can reduce the minimum distance for $3 \times 3$ kernels.

% For benchmarking, we plot the $\order(N^{-2})$ approximation to the Polyanskiy-Poor-Verd\'{u}'s metaconverse lower bound on the optimal BLER from~\cite[Fig.~6]{Ers_IT_16} implemented in~\cite{spectre_github}.

The complexity of the SCOS decoder can be measured approximately using the `average number of node visits' (${\rm ANV})$, where ${\rm ANV}=1$ indicates a complexity similar to that of the SC decoder~\cite{YuC_TCOM_24}. At $E_b/N_0=2.0$~dB, the value of ${\rm ANV}$ for  ${\rm d}\bid(5,2,2)$, ${\rm d}\RM(8,2)$ and the $[256,42]$ CRC-aided Polar code are $24.3 \times 10^3$, $369$ and $5.7$, respectively.

\subsection{Binary Erasure Channel (BEC)}

% We present two groups of results. In the first we 
Fig.~\ref{fig:Rnorm_BEC}
shows the normalized rate $R_{\rm norm}$~\cite[equation~(299)]{PPV_IT_10} of all BiD codes (of length $243$) and all RM codes (of length $256$) at ${\rm BLER}=10^{-3}$ with and without dynamic freezing. 
We computed $R_{\rm norm}$ for each code using the special converse~\cite[Theorem~38]{PPV_IT_10} implemented in~\cite{spectre_github}. 
% The results are shown as scatter plots of $R_{\rm norm}$ versus rate in Fig.~\ref{fig:Rnorm_without_dfcons} and~\ref{fig:Rnorm_with_dfcons}. 
For dynamic freezing of BiD codes we use the kernel $\p{A}_3$ if $r_1 \leq 2$ and the kernel $\p{A}'_3$ if $r_1 \geq 3$; this choice of kernels gives the largest $R_{\rm norm}$.

% In the second group of results for the erasure channel we present three sets of waterfall curves of BLER versus the gap to capacity $\varepsilon - (1-R)$ where $\varepsilon$ is the channel erasure probability and $R$ is the rate of the code under consideration. 
% We use $\varepsilon - (1-R)$ instead of $\varepsilon$ as the horizontal axis to account for the differences in the rates of the codes being compared.
% Fig.~\ref{fig:522},~\ref{fig:523} and~\ref{fig:534} 

Fig.~\ref{fig:534} shows BLER versus the gap to channel capacity $\varepsilon - (1 - R) = R - (1 - \varepsilon)$, where $\varepsilon$ is the BEC erasure probability and $R$ is the code rate (to account for the differences in the code rates) for: 
% We use $\varepsilon - (1-R)$ instead of $\varepsilon$ as the horizontal axis to account for the differences in the rates of the codes.
% Fig.~\ref{fig:534} shows the BLER curves of:
% \begin{inparaenum}[(i)]% [label=(\roman*)]
% \item 
\emph{(i)}~$\bid(5,3,4)$;
% \item 
\emph{(ii)}~the RM code with rate closest to $\bid(5,3,4)$;
% \item 
\emph{(iii)}~the RM-Polar code~\cite{LST_arXiv_14} with rate equal to the BiD code;
% \item 
\emph{(iv)}~variants of these three codes with dynamic freezing; and 
% \item 
\emph{(v)}~lower bound on BLER~\cite[Theorem~38]{PPV_IT_10} computed for the rate and blocklength of $\bid(5,3,4)$ and ${\rm d}\bid(5,3,4)$. % and~\cite{spectre_github} computed for the rate and blocklength of the BiD code.
% \end{inparaenum}
Appendix~\ref{app:bec_simulations} compares the BLER of $\bid(5,2,2)$ and $\bid(5,2,3)$ with codes constructed using $\p{A}_2$.

In all these results (BI-AWGN and BEC) BiD and dBiD codes offer a competitive BLER (under ML decoding) compared to codes constructed using the $2 \times 2$ kernel. %, and dBiD codes have BLER close to the finite-length information-theoretic limits.

\section{Discussion}

While the SCOS decoder~\cite{YuC_TCOM_24} works with BiD codes, the complexity is significantly higher than RM codes. 
Efficient near-ML decoders are needed for BiD codes.
% It is necessary to investigate efficient decoders for BiD codes.
Dynamic freezing does not deteriorate the weight distribution of codes from $2 \times 2$ kernel~\cite{LZG_arXiv_19,LZLWYM_ISIT_21}. However, 
the minimum distance of ${\rm d}\bid(5,2,2)$ is smaller than that of $\bid(5,2,2)$.
% dynamic freezing reduces the minimum distance of $\bid(5,2,2)$ from $48$ to at the most $27$. 
We need pre-transformation techniques for BiD codes that preserve the minimum distance.

% % % % % % % % % % % % % % % % % % % % 
% Bibliography
% \newpage 
% \bibliographystyle{IEEEtran}
\IEEEtriggeratref{11}
% \bibliography{IEEEabrv,ITW25} 


% % % % % % % % % % % % % % % % % % % % 

\newpage 

\appendix 

\subsection{Proof of Lemma~\ref{lem:G_mw_recursion}} 
\label{app:lem:G_mw_recursion}

For $N=3^m$ every row of $\p{A}_{N}$ is an $m$-fold Kronecker product $\p{a}_1 \otimes \cdots \otimes \p{a}_{m}$, where $\p{a}_1,\dots,\p{a}_m$ are length-$3$ vectors chosen from among the rows of $\p{A}_3$. 
Since $\p{A}_3$ has three distinct rows, we arrive at $3^m$ possibilities for $\p{a}_1 \otimes \cdots \otimes \p{a}_{m}$, exactly one per each row of $\p{A}_{N}$. Note that 
\begin{align*}
\wt\left( \p{a}_1 \otimes \cdots \otimes \p{a}_{m} \right) = \prod_{i=1}^{m} \wt(\p{a}_i).
\end{align*}
Since the first row $(1,1,1)$ of $\p{A}_3$ has weight $3$, and the remaining two rows have weight $2$, 
\begin{align*}
\wt\left( \p{a}_1 \otimes \cdots \otimes \p{a}_{m} \right) = 2^{w} \times 3^{m-w}
\end{align*}
where $w= \left| \{ i : \p{a}_i \neq (1,1,1)\} \right|$. 

Now consider the case $1 \leq w \leq m-1$. The matrix $\p{G}_{m,w}$ contains all rows of $\p{A}_{N}$ with weight $2^w \times 3^{m-1}$. Consider any such row $\p{a}_1 \otimes \cdots \otimes \p{a}_m$. 
If $\p{a}_1=(1,1,1)$, then exactly $w$ components of $\p{a}_2 \otimes \cdots \otimes \p{a}_m$ are either $(1,1,0)$ or $(1,0,1)$. Hence, $\p{a}_1 \otimes \cdots \otimes \p{a}_m$ is one of the rows of the matrix $(1,1,1) \otimes \p{G}_{m-1,w}$. 
If $\p{a}_1=(1,1,0)$, then $w-1$ components of $\p{a}_2 \otimes \cdots \otimes \p{a}_m$ are either $(1,1,0)$ or $(1,0,1)$. Hence, $\p{a}_1 \otimes \cdots \otimes \p{a}_m$ is a row of $(1,1,0) \otimes \p{G}_{m-1,w-1}$. 
A similar result holds for the case $\p{a}_1=(1,0,1)$. 
Gathering all the rows of $\p{G}_{m,w}$ according to the value of their first component $\p{a}_1$ gives us the result in~\eqref{eq:G_mw_matrix_recursion}.

The proofs for the cases $w=0$ and $w=m$ are similar.

\subsection{Proof of Theorem~\ref{thm:single_w_gen_matrix}} \label{app:thm:single_w_gen_matrix}

% \begin{IEEEproof}
% Recall that $\supp(\p{\rho}_{m,\p{j}}) = \{ \p{i} \in \Zb_3^m : \p{i} \cdot \p{j} \neq 1 \}$.
We will prove the theorem for $1 \leq w \leq m-1$. The proof is similar for $w=0$ (the repetition code) and $w=m$ (the $m$-fold product of the $[3,2,2]$ single-parity check code).
Assume $w \in \{1,\dots,m-1\}$.
The rows of the inverse-DFT-based generator matrix are indexed using $\{ \p{j} \in \Zb_3^m : \wt(\p{j}) = w\}$. 
The columns are indexed by $\p{i} = (i_1,\dots,i_m) \in \Zb_3^m$. Let us arrange the columns in the increasing order of $\sum_{\ell=1}^{m} i_{\ell} \, 3^{\ell-1}$.
Denote the $\p{j}^\tth$ row (for the length-$3^m$ code) as $\p{\rho}_{m,\p{j}}$, that is, 
\begin{equation*}
\supp(\p{\rho}_{m,\p{j}}) = \{ \p{i} \in \Zb_3^m : \p{i} \cdot \p{j} \neq 1 \}.
\end{equation*}
We need to show that \mbox{$\p{\rho}_{m,\p{j}} \in \rowspaceof(\p{G}_{m,w})$} for \mbox{$w=\wt(\p{j})$}. 

The case $m=1$ can be verified directly. We now prove the induction step.
Let $\p{j}=(j_1,\dots,j_m)$. For any $\p{i}$ and $\p{j}$ define the $(m-1)$-tuples $\p{i}'=(i_1,\dots,i_{m-1})$ and $\p{j}' = (j_1,\dots,j_{m-1})$.

\emph{Case~1: $j_m=0$.} We have $\p{i} \cdot \p{j} = \sum_{\ell=1}^{m-1} i_{\ell} j_{\ell} = \p{i}' \cdot \p{j}'$.
Since the entries of $\p{\rho}_{m,\p{j}}$ are independent of $i_m$ and $j_m$ we have 
\begin{equation*}
\p{\rho}_{m,\p{j}} = 
\left(
\p{\rho}_{m-1,\p{j}'}, \p{\rho}_{m-1,\p{j}'},  \p{\rho}_{m-1,\p{j}'}
\right)
= (1,1,1) \otimes \p{\rho}_{m-1,\p{j'}}.
\end{equation*}
Since \mbox{$\wt(\p{j}')=\wt(\p{j})=w$}, by the induction hypothesis we have $\p{\rho}_{m-1,\p{j}'} \in \rowspaceof(\p{G}_{m-1},w)$, thus, $\p{\rho}_{m,\p{j}} \in \rowspaceof\left( (1,1,1) \otimes \p{G}_{m-1,w} \right)$. 
To complete the proof of this case use the fact that $(1,1,1) \otimes \p{G}_{m-1,w}$ is a submatrix of $\p{G}_{m,w}$, see~\eqref{eq:G_mw_matrix_recursion}.

\emph{Case 2: \mbox{$j_m=1$}.} Observe that $\p{i}\cdot\p{j} = \p{i}' \cdot \p{j}' + i_m$. Let us partition $\p{\rho}_{m,\p{j}}$ into three subvectors of length $3^{m-1}$ each $\p{\rho}_{m,\p{j}} = (\p{a}_0,\p{a}_1,\p{a}_2)$. 
For $\ell \in \{0,1,2\}$, $\p{a}_\ell$ is the subvector corresponding to the column indices $\{ \p{i} \in \Zb_3^m : i_m = \ell \}$. Using $\p{i}'$ (the first $(m-1)$ components of $\p{i}$) as an index for the entries of $\p{a}_{\ell}$ we have
\begin{equation*}
\supp(\p{a}_{\ell}) = \left\{ \p{i}' \in \Zb_3^{m-1} : \p{i}' \cdot \p{j}' + \ell \neq 1 \right\}.
\end{equation*}
The support of $\p{a}_0$ coincides with the definition of the support of $\p{\rho}_{m-1,\p{j}'}$, hence, $\p{a}_0 = \p{\rho}_{m-1,\p{j}'}$. 
Now, for $\p{a}_2$ we have 
\begin{align*}
\supp(\p{a}_2) &= 
\left\{
\p{i}' : \p{i}' \cdot \p{j}' \neq 2
\right\} \\
&= 
\left\{
\p{i}' : \p{i}' \cdot 2\p{j}' \neq 1
\right\} ~~~~~~~(\text{since } 2^2=1 \text{ in } \Zb_3)\\
&= \supp(\p{\rho}_{m-1,2\p{j}'}).
\end{align*}
Hence \mbox{$\p{a}_2 = \p{\rho}_{m-1,2\p{j}'}$}. Note that, since $2$ is invertible in $\Zb_3$, the Hamming weights of $\p{j}'$ and $2\p{j}'$ are equal.
Finally for $\p{a}_1$, using $\triangle$ to denote the symmetric difference between two sets, 
% we see that 
\begin{align*}
\supp(\p{a}_1) &= \{\p{i}' : \p{i}' \cdot \p{j}' \neq 0\} \\
&= \{\p{i}' : \p{i}' \cdot \p{j}' \neq 1 \} ~\triangle~ \{\p{i}' : \p{i}' \cdot \p{j}' \neq 2 \} \\
&= \supp(\p{a}_0) ~\triangle~ \supp(\p{a}_2) \\
&= \supp(\p{\rho}_{m-1,\p{j}'}) ~\triangle~ \supp(\p{\rho}_{m-1,2\p{j}'}).
\end{align*}
Hence, $\p{a}_1 = \p{\rho}_{m-1,\p{j}'} + \p{\rho}_{m-1,2\p{j}'}$. Putting these results together 
% we see that 
\begin{align*}
\p{\rho}_{m,\p{j}} =& (1,1,0) \otimes \p{\rho}_{m-1,\p{j}'} + (0,1,1) \otimes \p{\rho}_{m-1,2\p{j}'} \\
=& (1,1,0) \otimes \p{\rho}_{m-1,\p{j}'} + (1,1,0) \otimes \p{\rho}_{m-1,2\p{j}'} \\
&~~~~~~~~~~~~+(1,0,1) \otimes \p{\rho}_{m-1,2\p{j}'}.
\end{align*}
Since $j_m=1$, $\wt(\p{j}') = \wt(2\p{j}') = \wt(\p{j})-1 = w-1$. By induction hypothesis $\p{\rho}_{m-1,\p{j}'}, \, \p{\rho}_{m-1,2\p{j}'} \in \rowspaceof(\p{G}_{m-1,w-1})$. 
Using this result with~\eqref{eq:G_mw_matrix_recursion} completes the proof for the case $j_m=1$.

\emph{Case 3: $j_m=2$.} This is similar to Case~2.
% \end{IEEEproof}

\subsection{Abelian Codes with Odd \& Even Frequency Weight Sets} \label{app:lem:odd_even_weight_set_dmin}

In contrast to the frequency weight sets of BiD codes, which are collections of contiguous integers, consider frequency weight sets that contain every other integer, 
\begin{align*}
\Wc_{m,e} &= \{ r \in \{0,\dots,m\}: r \text{ even}\}, \text{ and } \\
\Wc_{m,o} &= \{ r \in \{0,\dots,m\}: r \text{ odd}\}.
\end{align*}
The codes corresponding to $\Wc_{m,e}$ and $\Wc_{m,o}$ are a dual pair, and the sum of the codes is $\Fb_2^{3^m}$. The rates of these codes are close to $0.5$. 
Unlike BiD codes, the minimum distance of these codes is logarithmic in $N$, and hence, $\log \left( d_{\min} \right) / \log (N) \to 0$. 
To prove this result, we rely on the recursion~\eqref{eq:G_mw_matrix_recursion} and inductively identify a non-zero codeword of weight $\order(\log N)$.
In spite of this slow growth of $d_{\min}$, these codes have impressive EXIT functions in the binary erasure channel (see~\cite[Fig.~2]{NaK_IT_23}).

% \begin{lemma} \label{lem:odd_even_weight_set_dmin}
We now show that the minimum distances of $\abelian(m,\Wc_e)$ and $\abelian(m,\Wc_o)$ are at the most $2m+1$ and $2m$, respectively, which are both $\order(\log N)$.
% \end{lemma}
% \begin{IEEEproof}
% Please see Appendix~\ref{app:lem:odd_even_weight_set_dmin}.
% \end{IEEEproof}
% The following observation will help in bounding the minimum distance.
% \begin{lemma} \label{lem:sum_of_rows_of_A_N}
We first make the observation that the sum of all the rows of $\p{A}_3^{\otimes m}$ is $\p{e}_1$ (the first standard basis vector) for any $m$.
% \end{lemma}
% \begin{IEEEproof}
To see this, let $\p{1}_{N}$ denote the all-one vector of length $N$. The sum of all the rows of $\p{A}_3^{\otimes m}$ is 
\begin{align*}
\p{1}_N \p{A}_3^{\otimes m} = \p{1}_3^{\otimes m} \p{A}_3^{\otimes m} = \left( \p{1}_3 \p{A}_3 \right)^{\otimes m} = \left( (1,0,0) \right)^{\otimes m} = \p{e}_1.
\end{align*}
% \end{IEEEproof}

Denote the generator matrices corresponding to the weight sets $\Wc_{m,e}$ and $\Wc_{m,o}$ as $\p{G}_{m,e}$ and $\p{G}_{m,o}$, respectively. The recursion~\eqref{eq:G_mw_matrix_recursion} applied to these weight sets implies 
\begin{align*}
\p{G}_{m,e} 
\!=\!
\left[ \!\!
\begin{array}{l}
    (1,1,1) \otimes \p{G}_{m-1,e} \\
    (1,1,0) \otimes \p{G}_{m-1,o} \\
    (1,0,1) \otimes \p{G}_{m-1,o}
\end{array}
\!\!\right]
\!, 
\p{G}_{m,o} 
\!=\!
\left[ \!\!
\begin{array}{l}
    (1,1,1) \otimes \p{G}_{m-1,o} \\
    (1,1,0) \otimes \p{G}_{m-1,e} \\
    (1,0,1) \otimes \p{G}_{m-1,e}
\end{array}
\!\!\right]
\!\!,
\end{align*}
with $\p{G}_{1,e}=[1~1~1]$ and $\p{G}_{1,o} = [1~1~0;~1~0~1]$.
Using $\p{1}$ to denote the all-one vector of appropriate dimension, the sum of all the rows of $\p{G}_{m,e}$ is 
\begin{align*}
\p{1}\p{G}_{m,e} &= 
(1,1,1) \otimes \p{1}\p{G}_{m-1,e} 
+ (1,1,0) \otimes \p{1}\p{G}_{m-1,o} \\
&~~~~~~~~~~~~~~~~~~+ (1,0,1) \otimes \p{1}\p{G}_{m-1,o} \\
&= (1,1,1) \otimes \p{1}\p{G}_{m-1,e} 
+ (0,1,1) \otimes \p{1}\p{G}_{m-1,o} \\
&\stackrel{\text{(a)}}{=} \left( \p{1}\p{G}_{m-1,e},\, \p{1}\p{A}_3^{\otimes (m-1)}, \,\p{1}\p{A}_3^{\otimes (m-1)} \right) \\
&\stackrel{\text{(b)}}{=} \left( \p{1}\p{G}_{m-1,e}, \, \p{e}_1, \, \p{e}_1 \right),
\end{align*}
where (a) follows from the fact $\p{A}_3^{\otimes m} = [\p{G}_{m,e}^T ~ \p{G}_{m,o}^T]^T$ up to a permutation of rows, and (b) uses the fact that the sum of all the rows of $\p{A}_3^{\otimes (m-1)}$ is $\p{e}_1$. 
Hence, the weight of $\p{1} \p{G}_{m,e}$ is two more than the weight of $\p{1} \p{G}_{m-1,e}$. Using $\p{1}\p{G}_{1,e}=(1,1,1)$, we see that $\wt(\p{1}\p{G}_{m,e})=2m+1$ for all $m$. A similar derivation leads us to $\wt(\p{1}\p{G}_{m,o}) = 2m$.

\subsection{Proof of Theorem~\ref{thm:dmin_recursions}} \label{app:thm:dmin_recursions}

\emph{Case 1: $\p{\rho}_0=\p{\rho}_1=\p{\rho}_2=\p{0}$.} 
This corresponds to $\p{\rho}=\p{0}$, and is not useful for minimum distance analysis.

\emph{Case 2: exactly one among $\p{\rho}_0,\p{\rho}_1,\p{\rho}_2$ is non-zero.} 
For the purpose of our analysis let $\p{\rho}_0 \neq \p{0}$ and $\p{\rho}_1=\p{\rho}_2=\p{0}$; the analysis of the other two possibilities is similar. 
% Since $\p{a}_{r_2}, \p{a}_{r_1-1}'$ lie in different subspaces, $\p{\rho}_1=\p{0}$ necessitates $\p{a}_{r_2}=\p{a}_{r_1-1}'=\p{0}$ (
Since the sum-form of each $\p{\rho}_i$ mirrors the direct-sum decomposition~\eqref{eq:direct_sum_decomp}, $\p{\rho}_1=\p{0}$ if and only if $\p{a}_k=\p{a}_k'$ for all $k \in \{r_1,\dots,r_2-1\}$ and $\p{a}_{r_2} = \p{a}'_{r_1-1} = \p{0}$.
The same holds for $\p{\rho}_2$, and hence, $\p{a}_k=\p{a}_k''$ for all $k \in \{r_1,\dots,r_2-1\}$ and $\p{a}''_{r_1-1} = \p{0}$.
Hence, $\p{\rho}_0 = \sum_{k=r_1}^{r_2-1} \p{a}_k''$. 
If $r_1=r_2$, i.e., if $|\Wc|=1$, $\p{\rho}_0=\p{0}$, so Case~2 leads to a contradiction and is invalid. Otherwise, we see that $\p{\rho}_0$ is a non-zero codeword in $\abelian(m-1,\Wc_{0,-1})$.
By allowing a free choice of $\p{a}_k'' \in \abelian(m,\{k\})$ for all $k\in\{r_1,\dots,r_2-1\}$, we see that $\p{\rho}_0$ can be equal to any non-zero codeword of $\abelian(m,\Wc_{0,-1})$.
Hence, the smallest Hamming weight of $\p{\rho}$ under Case~2 is $D_2=d_{m-1}(\Wc_{0,-1})$, and this case is valid when $|\Wc| > 1$.

\emph{Case 3: exactly two among $\p{\rho}_0,\p{\rho}_1,\p{\rho}_2$ are non-zero.}
Consider \mbox{$\p{\rho}_1=\p{0}$} and \mbox{$\p{\rho}_0, \p{\rho}_2 \neq \p{0}$} (analysis of the other possibilities is similar). Using the direct-sum form~\eqref{eq:rho_012} of $\p{\rho}_1$, we see that Case~3 occurs if and only if \mbox{$\p{a}_k = \p{a}_k'$} for all $k \in \{r_1,\dots,r_2-1\}$, and \mbox{$\p{a}_{r_2}=\p{a}'_{r_1-1}=\p{0}$}. 
Thus, \mbox{$\p{\rho}_0=\sum_{k=r_1-1}^{r_2-1}\p{a}''_k$} and $\p{\rho}_2=\p{a}''_{r_1-1} + \sum_{k=r_1}^{r_2-1}(\p{a}'_k + \p{a}''_k)$, both of which belong to $\abelian(m,\Wc_{-1,-1})$. 
So the Hamming weight of $\p{\rho}$ in this case is at least $2d_{m-1}(\Wc_{-1,-1})$. 
Further, choosing $\p{\rho}_0$ to be a minimum weight codeword of $\abelian(m-1,\Wc_{-1,-1})$ and $\p{a}'_k=\p{0}$ for all $k \in \{r_1,\dots,r_2-1\}$, we see that $\p{\rho}_2=\p{\rho}_0$ and $\p{\rho}$ has weight exactly $2d_{m-1}(\Wc_{-1,-1})$. 
Hence, the smallest weight of $\p{\rho}$ under Case~3 is $D_3=2d_{m-1}(\Wc_{-1,-1})$.

\emph{Case~4: $\p{\rho}_0,\p{\rho}_1,\p{\rho}_2 \neq \p{0}$.} 
We identify two lower bounds for this case, and utilize their maximum. We also identify one recursive upper bound on $d_m(\Wc)$. 
% For a general choice of \mbox{$0<r_1 \leq r_2 < m$}, we are not sure if there exist codewords that meet this lower bound on Hamming weight. Hence, Case~4 doesn't contribute to the recursion for upper bound.
\begin{enumerate}
\item[\emph{4a.}] Since each $\p{\rho}_i \in \abelian(m-1,\Wc_{-1,0})$, we have $\wt\left( \p{\rho} \right) \geq D_{4a}=3d_{m-1}(\Wc_{-1,0})$. Because of the dependencies in the decomposition~\eqref{eq:rho_012} of $\p{\rho}_0,\p{\rho}_1,\p{\rho}_2$, we do not know if equality is possible in general. 
In fact, when $r_1 + r_2 \leq m$, a repeated application of solely this bound yields $3^{m-r_2}$ as a lower bound for $\bid(m, r_1,r_2)$. This is the minimum distance of the dual Berman code $\bid(m,0,r_2)$ which contains $\bid(m,r_1,r_2)$ as a subcode.

\item[\emph{4b.}] For the alternative lower bound, consider two subcases: 
\emph{(i)}~$\p{\rho}_0,\p{\rho}_1,\p{\rho}_2$ are all equal, and 
\emph{(ii)}~at least two of them are unequal. 

Subcase~\emph{(i)} occurs if and only if $\p{a}'_{k}=\p{a}''_{k}=\p{0}$ for all $k \in \{r_1-1,\dots,r_2-1\}$, and thus, $\p{\rho}_i = \sum_{k \in \Wc}\p{a}_k \in \abelian(m-1,\Wc)$ for all $i$. Hence, the smallest Hamming weight under subcase~\emph{(i)} is $D'_4 = 3d_{m-1}(\Wc)$. This subcase contributes to the recursion for both the upper and the lower bound on $d_m(\Wc)$.

Finally, consider subcase~\emph{(ii)}.
For concreteness, assume $\p{\rho}_0 \neq \p{\rho}_2$ (other possibilities can be handled in a similar way). 
From triangle inequality, 
\begin{align*}
\wt(\p{\rho}_0)+\wt(\p{\rho}_1)+\wt(\p{\rho}_2) \geq \wt(\p{\rho}_1)+\wt(\p{\rho}_0 + \p{\rho}_2).    
\end{align*} 
Using~\eqref{eq:rho_012}, we notice that $\p{\rho}_0 + \p{\rho}_2 \in \abelian(m-1,\Wc_{-1,-1})$ and $\p{\rho}_1 \in \abelian(m-1,\Wc_{-1,0})$. Thus, $d_{m}(\Wc) \geq d_{m-1}(\Wc_{-1,-1})+d_{m-1}(\Wc_{-1,0})$.
\end{enumerate}
Summarizing Case~4, when all $\p{\rho}_i$ are non-zero, the smallest value of $\wt(\p{\rho})$ is upper bounded by $D'_4=3d_{m-1}(\Wc)$, and is lower bounded by 
\begin{align*}
&\max\Big\{\, D_{4a}, \,\, \min\{D'_4,d_{m-1}(\Wc_{-1,-1})+d_{m-1}(\Wc_{-1,0})\} \, \Big\} \\
=& \max\{D_{4a},D_{4b}\}\\ =& D_4.
\end{align*}
Putting together the bounds identified in all the cases gives us the promised recursions for upper and lower bounds on $d_m(\Wc)$.

\subsection{List of all BiD Codes of Lengths $9$ to $729$}
\label{app:list_of_bid_codes}

\renewcommand{\arraystretch}{1.10}
\begin{table}[htbp]
    \centering
    \caption{All BiD Codes of Lengths $9$ to $243$}
    \label{table:bid_9_to_243}
\begin{tabular}{|c|c|c|c|c|}
    \hline
    \hline
    $m$ & $r_1$ & $r_2$ & $\dmin$& $K$ \\
    \hline
    \hline
        % 0 & 0 & 0 & 1 & 1 \\ \hline
        % 1 & 0 & 0 & 3 & 1 \\ \hline
        % 1 & 0 & 1 & 1 & 3 \\ \hline
        % 1 & 1 & 1 & 2 & 2 \\ \hline
        \multicolumn{5}{c}{Codes of length $9$} \\ \hline 
        2 & 0 & 0 & 9 & 1 \\ \hline
        2 & 0 & 1 & 3 & 5 \\ \hline
        2 & 0 & 2 & 1 & 9 \\ \hline
        2 & 1 & 1 & 4 & 4 \\ \hline
        2 & 1 & 2 & 2 & 8 \\ \hline
        2 & 2 & 2 & 4 & 4 \\ \hline \hline 
        \multicolumn{5}{c}{Codes of length $27$} \\ \hline 
        3 & 0 & 0 & 27 & 1 \\ \hline
        3 & 0 & 1 & 9 & 7 \\ \hline
        3 & 0 & 2 & 3 & 19 \\ \hline
        3 & 0 & 3 & 1 & 27 \\ \hline
        3 & 1 & 1 & 12 & 6 \\ \hline %
        3 & 1 & 2 & 4 & 18 \\ \hline %
        3 & 1 & 3 & 2 & 26 \\ \hline %
        3 & 2 & 2 & 6 & 12 \\ \hline
        3 & 2 & 3 & 4 & 20 \\ \hline %
        3 & 3 & 3 & 8 & 8 \\ \hline \hline
        \multicolumn{5}{c}{Codes of length $81$} \\ \hline 
        4 & 0 & 0 & 81 & 1 \\ \hline %
        4 & 0 & 1 & 27 & 9 \\ \hline
        4 & 0 & 2 & 9 & 33 \\ \hline
        4 & 0 & 3 & 3 & 65 \\ \hline
        4 & 0 & 4 & 1 & 81 \\ \hline %
        4 & 1 & 1 & 36 & 8 \\ \hline
        4 & 1 & 2 & 12 & 32 \\ \hline
        4 & 1 & 3 & 4 & 64 \\ \hline
        4 & 1 & 4 & 2 & 80 \\ \hline %
        4 & 2 & 2 & 16-18 & 24 \\ \hline
        4 & 2 & 3 & 6 & 56 \\ \hline
        4 & 2 & 4 & 4 & 72 \\ \hline %
        4 & 3 & 3 & 12 & 32 \\ \hline
        4 & 3 & 4 & 8 & 48 \\ \hline
        4 & 4 & 4 & 16 & 16 \\ \hline \hline 
        \multicolumn{5}{c}{Codes of length $243$} \\ \hline 
        5 & 0 & 0 & 243 & 1 \\ \hline
        5 & 0 & 1 & 81 & 11 \\ \hline
        5 & 0 & 2 & 27 & 51 \\ \hline
        5 & 0 & 3 & 9 & 131 \\ \hline
        5 & 0 & 4 & 3 & 211 \\ \hline
        5 & 0 & 5 & 1 & 243 \\ \hline
        5 & 1 & 1 & 108 & 10 \\ \hline
        5 & 1 & 2 & 36 & 50 \\ \hline
        5 & 1 & 3 & 12 & 130 \\ \hline
        5 & 1 & 4 & 4 & 210 \\ \hline
        5 & 1 & 5 & 2 & 242 \\ \hline
        5 & 2 & 2 & 48-54 & 40 \\ \hline
        5 & 2 & 3 & 16-18 & 120 \\ \hline
        5 & 2 & 4 & 6 & 200 \\ \hline
        5 & 2 & 5 & 4 & 232 \\ \hline
        5 & 3 & 3 & 22-36 & 80 \\ \hline
        5 & 3 & 4 & 12 & 160 \\ \hline
        5 & 3 & 5 & 8 & 192 \\ \hline
        5 & 4 & 4 & 24 & 80 \\ \hline
        5 & 4 & 5 & 16 & 112 \\ \hline
        5 & 5 & 5 & 32 & 32 \\ \hline 
    \end{tabular}
\end{table}

\renewcommand{\arraystretch}{1.10}
\begin{table}[htbp]
    \centering
    \caption{All BiD Codes of Length $729$}
    \label{table:bid_729}
\begin{tabular}{|c|c|c|c|c|}
    \hline
    \hline
    $m$ & $r_1$ & $r_2$ & $\dmin$& $K$ \\
    \hline
    \hline
        6 & 0 & 0 & 729 & 1 \\ \hline
        6 & 0 & 1 & 243 & 13 \\ \hline
        6 & 0 & 2 & 81 & 73 \\ \hline
        6 & 0 & 3 & 27 & 233 \\ \hline
        6 & 0 & 4 & 9 & 473 \\ \hline
        6 & 0 & 5 & 3 & 665 \\ \hline
        6 & 0 & 6 & 1 & 729 \\ \hline
        6 & 1 & 1 & 324 & 12 \\ \hline
        6 & 1 & 2 & 108 & 72 \\ \hline
        6 & 1 & 3 & 36 & 232 \\ \hline
        6 & 1 & 4 & 12 & 472 \\ \hline
        6 & 1 & 5 & 4 & 664 \\ \hline
        6 & 1 & 6 & 2 & 728 \\ \hline
        6 & 2 & 2 & 144-162 & 60 \\ \hline
        6 & 2 & 3 & 48-54 & 220 \\ \hline
        6 & 2 & 4 & 16-18 & 460 \\ \hline
        6 & 2 & 5 & 6 & 652 \\ \hline
        6 & 2 & 6 & 4 & 716 \\ \hline
        6 & 3 & 3 & 64-108 & 160 \\ \hline
        6 & 3 & 4 & 22-36 & 400 \\ \hline
        6 & 3 & 5 & 12 & 592 \\ \hline
        6 & 3 & 6 & 8 & 656 \\ \hline
        6 & 4 & 4 & 36-72 & 240 \\ \hline
        6 & 4 & 5 & 24 & 432 \\ \hline
        6 & 4 & 6 & 16 & 496 \\ \hline
        6 & 5 & 5 & 48 & 192 \\ \hline
        6 & 5 & 6 & 32 & 256 \\ \hline
        6 & 6 & 6 & 64 & 64 \\ \hline
    \end{tabular}
\end{table}

We present the code dimension (denoted as $K$) and the bounds on the minimum distance ($\dmin$) of all BiD codes of lengths $3^2=9$ to $3^6=729$. 
The upper and lower bounds on the minimum distance are computed recursively using Theorem~\ref{thm:dmin_recursions}. 
If the computed upper and lower bounds are not equal we present both these numbers, otherwise we present the exact value of $\dmin$.
Table~\ref{table:bid_9_to_243} shows BiD codes of lengths $9$ to $243$, while Table~\ref{table:bid_729} lists codes of length $729$.
Note that our simulations show that the minimum distance of $\bid(5,2,2)$ is equal to $48$.

\subsection{Proof of Theorem~\ref{thm:bid_dmin_thoeretical_lower_bound}}
\label{app:thm:bid_dmin_theoretical_lower_bound}

% We only provide the outline of the proof technique here.
We need to show that each of the two expressions $4^{r_1} \times 3^{m-r_1-r_2}$ and $3^{m-r_2} \times 2^{r_1 + r_2 - m}$ is a valid lower bound. 
This can be shown via induction using the recursion structure of Theorem~\ref{thm:dmin_recursions}. 
For induction, assume that the statement is true for all BiD codes of length $3^{m-1}$ and use the lower bound in Theorem~\ref{thm:dmin_recursions}; this is straightforward. 
The recursion terminates at weight sets corresponding to Berman and dual Berman codes. It is easy to verify that both the expressions are valid lower bounds on $\dmin$ for Berman and dual Berman codes of all lengths.

We will show that $3^{m-r_2} \times 2^{r_1 + r_2 - m}$ is a lower bound on the minimum distance of $\abelian(m,\{r_1,\dots,r_2\})$; the proof of the other expression follows similar steps. 
First, we verify if the base cases (Berman codes and their duals) of the recursion are satisfied by the expression. 
The minimum distance of the Berman code $\abelian(m,\{r_1,\dots,m\})$ is $2^{r_1}$ which is equal to the expression when $r_2=m$. The minimum distance of the dual Berman code $\abelian(m,\{0,\dots,r_2\})$ is $3^{m-r_2}$, and this is indeed lower bounded by the expression when $r_1=0$, i.e., by $(3/2)^{m-r_2}$.

To prove the induction step, we assume that the statement of the theorem is true for all BiD codes of length $3^{m-1}$, and prove the result for codes of length $3^m$. We now compute lower bounds on $D_2,D_3,D_4$. 

Note that $D_2$ is the minimum distance of the abelian code $\abelian(m-1,\{r_1,\dots,r_2-1\})$. By induction hypothesis, this is at least $3^{m-1-(r_2-1)} \times 2^{r_1 + r_2 - 1 - (m-1)} = 3^{m-r_2} \times 2^{r_1 + r_2 - m}$.

For $D_3$, note that $\Wc_{-1,-1}$ is $\{r_1-1,\dots,r_2-1\}$. Hence, $D_3$ is lower bounded by $2 \times 3^{m-1-(r_2-1)} \times 2^{r_1-1+r_2-1+m-1}$, which equals $3^{m-r_2} \times 2^{r_1 + r_2 - m}$.

Now consider $D_{4a}$. To lower bound this, we need to consider $\Wc_{-1,0}=\{r_1-1,\dots,r_2\}$ as a frequency weight set of code of length $3^{m-1}$. From the induction hypothesis, $D_{4a}$ is at least $3 \times 3^{m-1-r_2} \times 2^{r_1-1+r_2-(m-1)}$, which equals $3^{m-r_2} \times 2^{r_1 + r_2 - m}$. Note that $D_4 \geq D_{4a}$. 

Hence, $\min\{D_2,D_3,D_4\} \geq 3^{m-r_2} \times 2^{r_1 + r_2 - m}$.

\subsection{Proof of Theorem~\ref{thm:W_equal_to_1}} 
\label{app:thm:W_equal_to_1}

% We first consider $\Wc=\{1\}$. 
Let $\bar{d}_{m}(\Wc)$ denote the maximum weight among all codewords in $\abelian(m,\Wc)$, and let $d_m(\Wc) = \dmin\left( \abelian(m,\Wc) \right)$. 
Using~\eqref{eq:G_mw_matrix_recursion} with $w=1$ and the fact that $\abelian(m-1,\{0\})$ is the repetition code of length $3^{m-1}$, note that
% First note that using the generator matrix (from \ref{eq:G_mk}), 
an arbitrary codeword $\p{\rho}$ in the code $\bid(m,1,1)$ can be written as
\begin{align*}
%\label{eq:gen_1}
    \p{\rho} = \left( \, \p{a}_{1} + b\,\p{1} + c\,\p{1},~ \p{a}_{1} + b\,\p{1},~ \p{a}_{1} + c\,\p{1} \, \right)
    % \hat{\p{a}}_{m,\{1\}} = 
    % \left[
    % \begin{array}{l|l|l} 
    %     (a+b)\p{1} & (a)\p{1} & (b)\p{1} \\
    %     \p{a}_{m-1, \{1\}} & \p{a}_{m-1, \{1\}} & \p{a}_{m-1, \{1\}}
    % \end{array}
    % \right]
\end{align*}
where $\p{1}$ is the all-one vector of length $3^{m-1}$, $\p{a}_1 \in \abelian(m-1,\{1\}) = \bid(m-1,1,1)$, and $b,c \in \{0,1\}$. 
% The codeword generated is the sum of the two rows in \ref{eq:gen_1}. Thus, we require precisely 
We consider four cases for an exhaustive analysis of codewords (corresponding to $b,c \in \{0, 1\}$), grouped into two as below.
\begin{enumerate}
\item[\emph{(i)}] \emph{Case: \mbox{$(b,c)=(0,0)$}.} 
In this case \mbox{$\p{\rho}=(\p{a}_1,\p{a}_1,\p{a}_1)$}. 
The smallest non-zero weight of $\p{\rho}$ is $3d_{m-1}(\{1\})$ and the largest is $3\bar{d}_{m-1}(\{1\})$. 

\item[\emph{(ii)}] \emph{Case: \mbox{$(b,c) \neq (0,0)$}.} The three possibilities of $(b,c)$ have similar analysis since for all of them $\wt\left( (b+c,b,c) \right)=2$. For \mbox{$b=c=1$}, we have $\p{\rho}=(\p{a}_1, \p{1} + \p{a}_1, \p{1} + \p{a}_1)$. 
Observe that $\wt(\p{1} + \p{a}_1) = 3^{m-1} - \wt(\p{a}_1)$.
Thus, $\wt(\p{\rho}) = 2 \times 3^{m-1} - \wt(\p{a}_1)$. The smallest Hamming weight in this case is $2 \times 3^{m-1} - \bar{d}_{m-1}(\{1\})$, and the largest is $2 \times 3^{m-1} - d_{m-1}(\{1\})$.

% \item $(a,b)=(0,1)\equiv(1,0)$. $d_{m}(\Wc_{0,0}) = 2\times3^{m-1}-({\bar{d}})_{m-1}(\Wc_{0,0})$ and $\bar{d_{m}(\Wc_{0,0})}=2\times3^{m-1}-{d}_{m-1}(\Wc_{0,0})$

% \item $(a,b)=(1,1)$. $d_{m}(\Wc_{0,0}) = 2\times3^{m-1}-({\bar{d}})_{m-1}(\Wc_{0,0})$ and $\bar{d_{m}}(\Wc_{0,0})=2\times3^{m-1}-{d}_{m-1}(\Wc_{0,0})$ \textcolor{red}{$=$?}\comm{fixed}, which is the same as the second case. 
% \lp{The $(d_{\max})_{m-1}$ notation is a little clumsy; use another another notation instead? like $\bar{d}_{m-1}(\Wc)$?}\comm{fixed}
\end{enumerate}
Putting together these cases we obtain the recursions
\begin{align*}
d_{m}(\{1\}) &= \min\{\,3d_{m-1}(\{1\}),~2 \times 3^{m-1} - \bar{d}_{m-1}(\{1\})\,\}, \\
\bar{d}_{m}(\{1\}) &= \max\{\,3\bar{d}_{m-1}(\{1\}),~2 \times 3^{m-1} - d_{m-1}(\{1\}) \,\}.
\end{align*}
An induction based argument that combines the above result with a direct verification for the case $m=2$ completes the proof.

\subsection{Proof of Theorem~\ref{thm:W_m_minus_1}} 
\label{app:thm:W_m_minus_1}

% We now consider the case $\Wc=\{m-1\}$. 
We show that $\dmin(\bid(m,m-1,m-1)) = 3 \times 2^{m-2}$ for all $m \geq 3$ using induction on $m$ via Theorem~\ref{thm:dmin_recursions}. 
We verified the induction base case \mbox{$m=3$} numerically. 
To prove the induction step for the code $\bid(m,m-1,m-1)=\abelian(m,\Wc)$ where $\Wc=\{m-1\}$, Theorem~\ref{thm:dmin_recursions} uses BiD codes of length $3^{m-1}$ with weight sets $\Wc_{-1,-1}=\{m-2\}$, $\Wc_{-1,0}=\{m-2,m-1\}$ and $\Wc=\{m-1\}$. The first code has minimum distance $3 \times 2^{m-3}$ based on induction hypothesis, and the latter two codes are Berman codes with minimum distances $2^{m-2}$ and $2^{m-1}$, respectively. Hence, the terms $D_3,D_4,D'_4$ in Theorem~\ref{thm:dmin_recursions} can be computed, and $D_2=+\infty$. 
Using these values, we observe that the upper and lower bounds from this theorem match, and are equal to $3 \times 2^{m-2}$. 

\subsection{Proof of Theorem~\ref{thm:bid_asymptotic_dmin}} \label{app:thm:bid_asymptotic_dmin}

Let $Z_1,\dots,Z_m$ be independent Bernoulli(2/3) random variables. The rate of $\bid(m,r_1,r_2)$ is 
\begin{align*}
& \frac{1}{3^m} \sum_{w=r_1}^{r_2} \binom{m}{w} 2^w = \sum_{w=r_1}^{r_2} \binom{m}{w} \left(\frac{2}{3}\right)^w \left( \frac{1}{3} \right)^{m-w} \\
=& \, \Pb \!\left[ r_1 \leq \sum_{i=1}^{m} Z_i \leq r_2 \right] \\
=& \, \Pb \left[ \frac{r_1 - 2m/3}{\sqrt{2m/9}} \leq \frac{\sum_{i=1}^{m} (Z_i - 2/3)}{\sqrt{2m/9}} \leq \frac{r_2-2m/3}{\sqrt{2m/9}}\right]
\end{align*}

Let $Q(t) = \left( 1/2\pi \right) \int_{t}^{\infty} e^{-x^2/2} {\rm d}x$.
Considering $m \to \infty$ and applying the central limit theorem, we see that choosing $r_1,r_2 \in \{0,\dots,m\}$ closest to 
\begin{align*}
\frac{2m}{3} - \sqrt{\frac{2m}{9}} Q^{-1}\left( \frac{1-R}{2} \right), 
\frac{2m}{3} + \sqrt{\frac{2m}{9}} Q^{-1}\left( \frac{1-R}{2} \right),
\end{align*}
respectively, gives us a sequence of BiD codes with rate converging to $R$. Observe that $r_1/m, r_2/m \to 2/3$ for this sequence of codes. 
From the second lower bound in Theorem~\ref{thm:bid_dmin_thoeretical_lower_bound} and using $N=3^m$, we have 
\begin{align*}
\frac{\log d_{\min}}{ \log N} \geq \frac{m-r_2}{m} + \left( \frac{r_1 + r_2 - m}{m} \right)\frac{\log 2}{\log 3}.
\end{align*}
The RHS in the above inequality converges to $\log(6) / \log(27)$ as $m \to \infty$. 

% \subsection{Proof of Lemma~\ref{lem:odd_even_weight_set_dmin}} 

\subsection{SCOS Decoding of BiD Codes} \label{app:decoding}

\begin{algorithm}[!t]
    \caption{${\rm recursivelyCalcL}(\lambda,\phi)$}
    \label{algo:recursivelyCalcL}
    % \vspace{5pt}
    \textbf{Input:} layer $\lambda$ and phase $\phi$ \\
    % \textbf{Output:} 
    \vspace{-5pt}
    \begin{algorithmic}[1]
    \If{$\lambda=1$}
    \State return 
    \EndIf
    \State $\psi = \lfloor \phi/3 \rfloor$, $t=3^{\lambda-2}$
    \If{$\phi \mod 3 = 0$}
    \State ${\rm recursiveCalcL}(\lambda-1,\psi)$
    \EndIf

    \For{$\beta=0,1,\dots,3^{\log_3 N - \lambda + 1}-1$}
    \If{$\phi \mod 3 = 0$} 
    \State \mbox{$L[\lambda, \phi + 3 \beta t + 1] =$} $f^{-}\left( L[\lambda-1, \psi + 3 \beta t + 1], L[\lambda-1, \psi + (3 \beta+2)t + 1] \right)$
    \ElsIf{$\phi \mod 3 = 1$}
    \State $t_1 = f^{+}(L[\lambda-1, \psi + 3 \beta t + 1], L[\lambda-1, \psi + (3 \beta+2)t + 1], C[\lambda,\phi+3 \beta t])$
    \State $t_2 = f^{+}(L[\lambda-1,\psi + (3 \beta+1)t + 1],0,C[\lambda,\phi+3 \beta t])$
    \State $L[\lambda, \phi + 3 \beta t + 1] = f^{-}(t_1,t_2)$
    \Else 
    \State $u_1 = (-1)^{C[\lambda, \phi+ 3 \beta t - 1]}$, $u_2 = (-1)^{C[\lambda,\phi+3 \beta t]}$
    \State $t_1 = L[\lambda-1,\psi + 3 \beta t + 1] \cdot u_1 u_2$
    \State $t_2 = L[\lambda-1, \psi + (3\beta + 1)t + 1] \cdot u_1$
    \State $t_3 = L[\lambda-1, \psi + (3 \beta + 2)t + 1] \cdot u_2$
    \State $L[\lambda, \phi + 3 \beta t + 1] = t_1 + t_2 + t_3$
    \EndIf
    \EndFor
    \end{algorithmic}
\end{algorithm}
% temp1 = f_plus(p.L[lambda-1][psi + l2*3*beta + 1], p.L[lambda-1][psi + l2*(3*beta+2) + 1], p.C[lambda][phi+l1*beta])
% temp2 = f_plus(p.L[lambda-1][psi + l2*(3*beta+1) + 1],0.0,p.C[lambda][phi+l1*beta])
% p.L[lambda][phi + l1*beta + 1] = f_minus(temp1,temp2) 

\begin{algorithm}[!t]
    \caption{${\rm recursivelyCalcC}(\lambda,\phi)$}
    \label{algo:recursivelyCalcC}
    % \vspace{5pt}
    \textbf{Input:} layer $\lambda$ and phase $\phi$ \\
    % \textbf{Output:} 
    \vspace{-5pt}
    \begin{algorithmic}[1]
    \State $\psi = \lfloor \phi/3 \rfloor$, $t=3^{\lambda-2}$
    \For{$\beta=0,1,\dots,3^{\log_3 N - \lambda + 1}-1$}
    \State $C[\lambda-1,\psi + 3\beta t + 1] = C[\lambda, \phi + 3\beta t - 1] \oplus C[\lambda, \phi + 3 \beta t] \oplus C[\lambda, \phi + 3 \beta t + 1]$
    \State $C[\lambda-1,\psi + (3\beta+1)t + 1] = C[\lambda, \phi + 3 \beta t - 1] \oplus C[\lambda, \phi + 3 \beta t + 1]$
    \State $C[\lambda-1, \psi + (3\beta+2)t + 1] = C[\lambda, \phi + 3 \beta t] \oplus C[\lambda, \phi + 3 \beta t + 1]$
    \EndFor

    \If{$\psi \mod 3 = 2$}
    \State ${\rm recursivelyCalcC}(\lambda-1,\psi)$
    \EndIf
    \end{algorithmic}
\end{algorithm}

\begin{figure}%[!t]
    \centering
    \includegraphics[width=\linewidth]{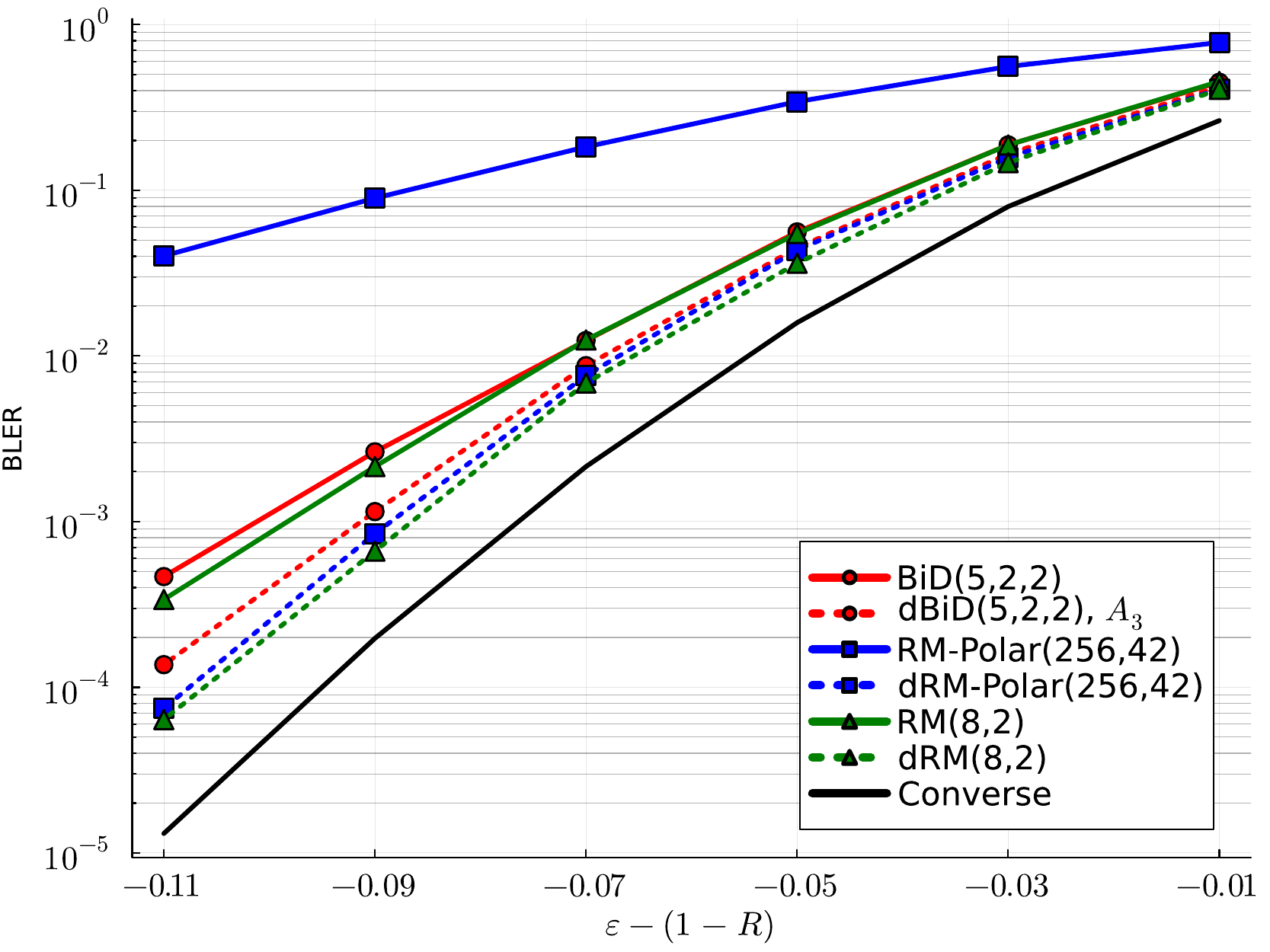}
    \caption{BLER of $\bid(5,2,2)$ and ${\rm d}\bid(5,2,2)$ (kernel $\p{A}_3$) in BEC.}
    \label{fig:522}
\end{figure}

\begin{figure}%[!t]
    \centering
        \includegraphics[width=\linewidth]{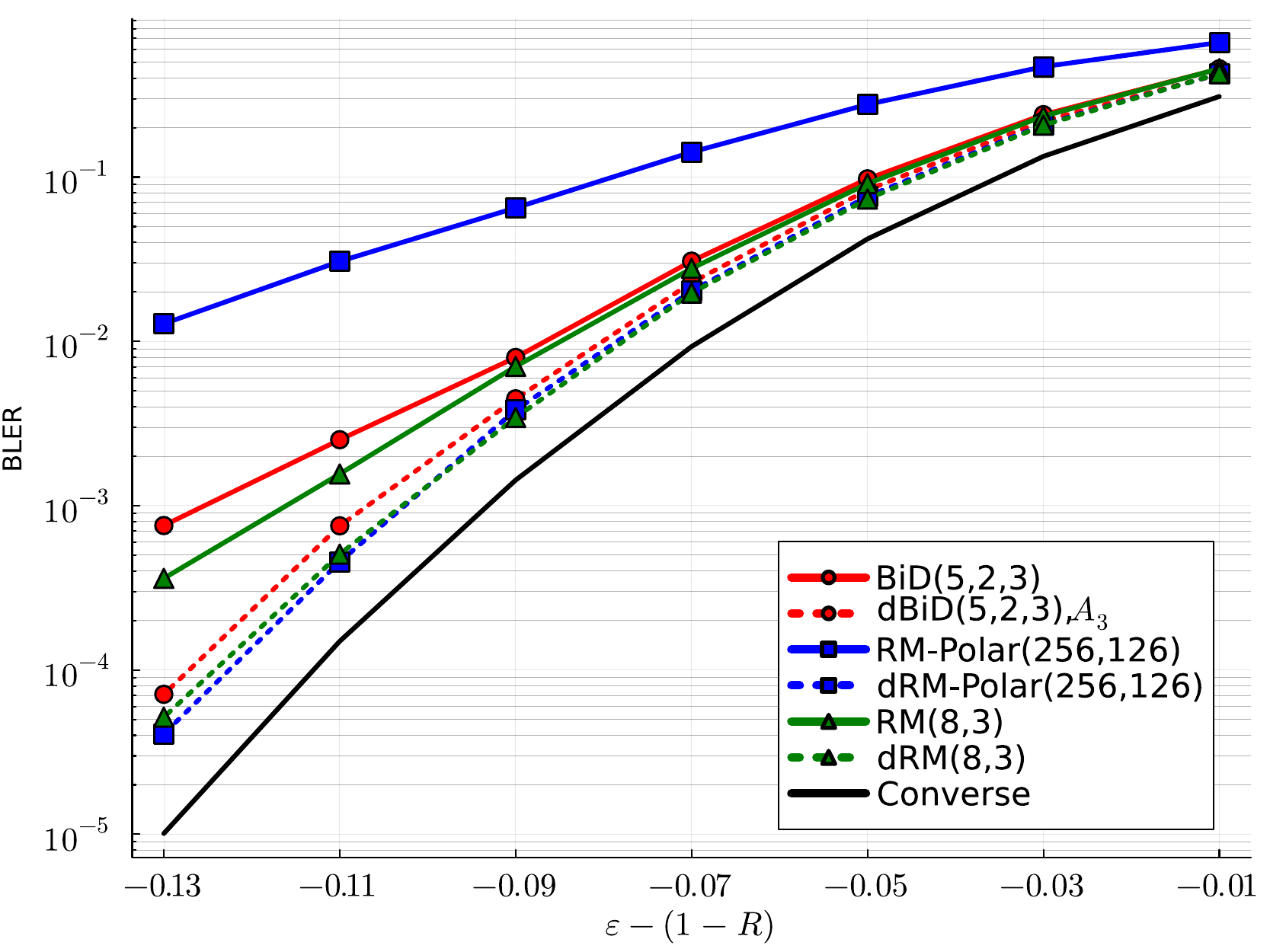}
    \caption{BLER of $\bid(5,2,3)$ and ${\rm d}\bid(5,2,3)$ (kernel $\p{A}_3$) in BEC.}
    \label{fig:523}
\end{figure}

The SCOS decoder was proposed for codes designed from Ar{\i}kan's kernel $\p{A}_2$ (including Polar codes of length $2^m$, RM codes, as well as codes with dynamic frozen bits and polarization-adjusted convolutional (PAC) codes~\cite{Arikan_PAC_2019,TrM_JSAC_16,YFV_Entropy_21}), but can be generalized to other kernels. 
We now highlight the main changes required to adapt the SCOS decoder to the kernel $\p{A}'_3$, which are along expected lines. 
In doing so, we borrow the notation from~\cite{YuC_TCOM_24} without introducing them a priori.
We request the reader to use~\cite{YuC_TCOM_24} as the primary reference for the treatment given in this subsection.

The main modifications are towards the functions ${\rm recursivelyCalcL}$ and ${\rm recursiveCalcC}$ used in SCOS. The versions adapted for the kernel $\p{A}'_3$ are shown in Algorithms~\ref{algo:recursivelyCalcL} and~\ref{algo:recursivelyCalcC}. Further, in the ${\rm SCDec}$ algorithm of~\cite{YuC_TCOM_24}, line~1 must be modified to `$m=\log_3 N$' and line~18 to `if $i \mod 3=0$'.

\subsection{Additional Simulation Results for BEC} \label{app:bec_simulations}

We present two more simulation results in Fig.~\ref{fig:522} 
% \comm{Figure 6 is on a different page?} 
and~\ref{fig:523} (in the same vein as Fig.~\ref{fig:534}) for the codes $\bid(5,2,2)$ and $\bid(5,2,3)$, respectively. As in Fig.~\ref{fig:534}, we compare these BiD codes (and their dynamically frozen variants) with RM codes (of closest rate as BiD codes) and the RM-Polar codes (with same rate as BiD codes), and the lower bound on BLER from~\cite{PPV_IT_10}. 

\hfill\IEEEQED

\end{document}

%% file: macros.tex
\newcommand{\suppress}[1]{}

\newcommand{\Pb}{\mathbbmss{P}} % probability
\newcommand{\Fb}{\mathbbmss{F}} % field
\newcommand{\Zb}{\mathbbmss{Z}} % integers
\newcommand{\wt}{{\rm w}_{\!H}}   % weight
\newcommand{\supp}{\mathsf{supp}}  % support set

\newcommand{\RM}{\mathrm{RM}}   % Reed-Muller code
\newcommand{\tth}{\text{th}}    % such as i^th --> $i^\tth$
\newcommand{\p}{\pmb}           % bold font
\newcommand{\dmin}{d_{\min}}    % min distance

\newcommand{\dimenof}{{\rm dim}}  % dimension
\newcommand{\spanof}{{\rm span}}  % span
\newcommand{\rowspaceof}{{\rm rowsp}}  % rowspaceofmatrix
\newcommand{\order}{\mathscr{O}}
\newcommand{\indicator}{\mathbbmss{1}}

\newcommand{\Wc}{\mathscr{W}}
\newcommand{\abelian}{\mathscr{C}_{\rm A}} % abelian code with (m,W)
\newcommand{\bid}{{\rm BiD}}